\documentclass{aa}  

\usepackage{graphicx}
\usepackage{txfonts}

\bibpunct{(}{)}{;}{a}{}{,} 

\newcommand{\rev}[1]{#1}

\usepackage{color}
\usepackage{hyperref}
\usepackage{tabularx}
\usepackage{subcaption}
\usepackage{graphicx}
\usepackage{placeins}

\makeatletter
\renewcommand*\aa@pageof{, page \thepage{} of \pageref*{LastPage}}
\makeatother

\hypersetup{%
        colorlinks,
        breaklinks=true,
        plainpages=false,%
        citecolor=[rgb]{0,0.20,0.45},
        linkcolor=[rgb]{0,0.20,0.45},
        urlcolor=[rgb]{0,0.20,0.45},
        bookmarksopen=true,%
        bookmarksnumbered=false,%
        bookmarksdepth=5%
}
\begin{document}

   \title{The spectral energy distribution of YSES 1 b\\ and its circumplanetary disc\thanks{This paper includes data gathered with the 6.5 meter Magellan Telescopes located at Las Campanas Observatory, Chile.}}

   \subtitle{}

   \author{Michiel Darcis \inst{1,2}\fnmsep\thanks{Corresponding author; m.darcis@sron.nl}
        \and 
        Sebastiaan Y. Haffert \inst{2,3}
        \and
        Tomas Stolker \inst{2}
        \and
        Richelle F. van Capelleveen \inst{2}
        \and
        Matthew A. Kenworthy \inst{2}
        \and
        Pieter J. de Visser \inst{1}
        \and
        Laird M. Close \inst{3}
        \and
        Olivier Guyon \inst{3, 4, 5, 6}
        \and
        Alexander D. Hedglen \inst{7}
        \and
        Parker T. Johnson \inst{3}
        \and
        Maggie Y. Kautz \inst{3}
        \and
        Jay K. Kueny \inst{5}
        \and
        Jialin Li \inst{3}
        \and
        Joseph D. Long \inst{8}
        \and
        Jennifer Lumbres \inst{3}
        \and
        Jared R. Males \inst{3}
        \and
        Eden A. McEwen \inst{5}
        \and
        Avalon L. McLeod \inst{9}
        \and
        Logan A. Pearce \inst{10}
        \and
        Lauren Schatz \inst{11}
        \and
        Kyle Van Gorkom \inst{3}
        }

   \institute{SRON - Space Research Organisation Netherlands, Niels Bohrweg 4, 2333 CA Leiden, The Netherlands
   \and
    Leiden Observatory, Leiden University, PO Box 9513, 2300 RA Leiden, The Netherlands
    \and
    Steward Observatory, The University of Arizona, 933 North Cherry Avenue, Tuscon, Arizona, USA
    \and
    Subaru Telescope, National Observatory of Japan, NINS, 650 N. A'ohoku Place, Hilo, Hawai'I, USA
    \and
    Wyant College of Optical Sciences, The University of Arizona, 1630 E University Boulevard, Tucson, Arizona, USA
    \and
    Astrobiology Center, National Institutes of Natural Sciences, 2-21-1 Osawa, Mitaka, Tokyo, Japan
    \and
    Northrop Grumman Corporation, 600 South Hicks Road, Rolling Meadows, Illinois, USA
    \and
    Center for Computational Astrophysics, Flatiron Institute, 162 5th Avenue, New York, New York, USA
    \and
    Draper Laboratory, 555 Technology Square, Cambridge, Massachusetts, USA
    \and
    Department of Astronomy, University of Michigan, 323 West Hall, 1085 S University Ave, Ann Arbor, MI 48109, USA
    \and
    Starfire Optical Range, Kirtland Air Force Base, Albuquerque, New Mexico, USA
    }

   \date{}

 
  \abstract
   {Direct imaging enables the characterisation of substellar companions on wide orbits.
   These objects provide a testbed for our formation theories; therefore, it is important to obtain accurate physical parameters for them.
   One of these objects is YSES 1 b.}
   {Our objective is to improve the spectral energy distribution (SED) modelling of YSES 1 b and determine the bulk and atmospheric parameters.}
   {We obtained observations in the $r'$, $i'$, and $z'$ bands using MagAO-X on the 6.5 metre Magellan Clay telescope at Las Campanas Observatory.
   We combined this data with archival VLT/SPHERE and VLT/NACO data and used a forward-modelling approach to estimate the physical parameters.
   We tested models both without and with a circumplanetary disc (CPD) model.
   We represented the CPD by including a dust extinction model and a blackbody radiation component.
   Using the derived bolometric luminosity, we estimated the mass of YSES 1 b by fitting evolutionary models.}
   {Including the CPD model provides a significantly better fit to the photometric data, yielding an object that is considerably warmer ($2854^{+110}_{-94}\,\mathrm{K}$ vs $1727^{+172}_{-127}\,\mathrm{K}$) and smaller ($1.58^{+0.06}_{-0.07}$ $\mathrm{R_J}$ vs $3.0^{+0.2}_{-0.7}\,\mathrm{R_J}$) than previous estimates.
   The newly determined radius suggests that the addition of dust extinction could resolve the large radius anomaly identified previously.
   Depending on the age of the system, the estimated mass increases from $14\pm3\,\mathrm{M_J}$ (17 Myr) to either $25.7^{+4.1}_{-3.6}$ (17 $\mathrm{Myr}$) or $41.6^{+3.6}_{-3.4}\,\mathrm{M_J}$ (27 $\mathrm{Myr}$).}
   {Dust extinction and blackbody radiation from a CPD can substantially change the estimated physical parameters of an object.
   For YSES 1 b, this \rev{moves it into} the brown dwarf regime.}

   
   \maketitle
%
\section{Introduction} \label{section:introduction}

Direct imaging of exoplanets provides valuable information on planet formation and demographics.
The sensitivity of current instrumentation enables the detection of young, giant, self-luminous planets on orbits with semi-major axes from tens to hundreds of $\mathrm{AU}$, which is a population of objects almost exclusively accessible through direct imaging \citep{currie_direct_2023}.
Furthermore, because photons are detected directly from the objects themselves, atmospheric characterisation is possible, which improves our physical understanding of these companions.

One of the systems that provides an interesting test for our planet formation theories is YSES 1, a young solar-type star located in the Lower Centaurus Crux subgroup of the Scorpius-Centaurus association \citep{pecaut_star_2016}.
Two substellar-companions (YSES 1 b and c) are known to orbit at projected separations of $160 \,\mathrm{AU}$ and $320 \,\mathrm{AU}$ \citep{bohn_young_2020, bohn_two_2020}.
Initial spectral energy distribution (SED) fits using photometric data from $1-5\,\mu\mathrm{m}$ provided mass estimates of $14\pm3\,M_J$ and $6\pm1\,M_J$ respectively.

Since its discovery, YSES 1 b has received considerable observational follow-up.
Abundance ratios have been measured to investigate the formation history. \citet{zhang_13co-rich_2021} first measured the $^{12}$CO/$^{13}$CO isotopologue ratio using medium-resolution $K$-band data from VLT/SINFONI 
and report an enriched value of $31^{+17}_{-10}$. 
A subsequent analysis based on VLT/CRIRES+ 
observations provides a different value of $88\pm13$ \citep{zhang_eso_2024}.
Despite the discrepancy between the two retrieved $^{12}$CO/$^{13}$CO ratios, both datasets suggest a solar-like C/O ratio.
More recently, \citep{roberts_new_2025} constrained the orbit of YSES 1 b using archival astrometry, radial velocity data, proper motion measurements, and new VLTI/GRAVITY 
data.
They find a moderate eccentricity of $0.44\pm0.20$, indicating a complex dynamical history.
They also revise the semi-major axis to a slightly smaller value of $146^{+16}_{-10}\,\mathrm{AU}$.

In addition to measuring abundance ratios, \citet{zhang_13co-rich_2021} detected emission from the Brackett $\gamma$ line, suggesting that the companion is still accreting from a circumplanetary disc (CPD).
\citet{holstein_survey_2021} attempted to detect the CPD using polarimetric observations but only \rev{found} an upper limit.
Another possible inference of the existence of a CPD is the relatively low projected rotational velocity measured in \citep{zhang_eso_2024}, \rev{where} magnetic braking from the CPD slows down the rotation.
More substantial evidence for the CPD recently came from \rev{James Webb Space Telescope (JWST)} spectroscopy\citep{hoch_silicate_2025}.
Based on the infrared excess in the SED, the best fit results in a blackbody disc temperature of approximately 500 K and a disc radius between $8.5-20\,\mathrm{R_J}$.
They explain a broad emission feature at $8-11\,\mu\mathrm{m}$ by sub-micron olivine grains.
Recently, \cite{julo_stellar_2025} used VLT/MUSE to measure the H$\alpha$, H$\beta$, CaII H\&K triplet, and HeI emission lines associated with accretion. From the H$\alpha$ line  they estimated an accretion rate of $\sim1.45\times10^{-9\pm0.19} \,\mathrm{M_J}$/year.
These findings place YSES 1 b within a small group of known substellar companions that host a CPD.
Other notable objects include GQ Lup b \citep{stolker_characterizing_2021}, PDS 70 c \citep{benisty_circumplanetary_2021, haffert_two_2019}, GSC 6214-210 b \citep{bowler_disk_2011, bailey_thermal_2013, bowler_alma_2015}, DH Tau b \citep{zhou_accretion_2014}, CT Cha b \citep{cugno_carbon-rich_2025}, and Delorme 1 (AB) b \citep{betti_near-infrared_2022, demars_exoplanet_2025}.

In this paper, we extend the photometric data of YSES 1 b to the optical bands using observations from the MagAO-X 
instrument \citep{males_magao-x_2024}.
Together with archival photometric data, we update the SED modelling of YSES 1 b and its CPD by including dust extinction and a blackbody emission component.
Section \ref{sec:observations} presents the MagAO-X observations and the data reduction to retrieve the contrast and astrometry.
Section \ref{sec:analysis} presents the results of our SED fits and mass estimates.
Section \ref{sec:discussion} discusses the implications of these results. 
Section \ref{sec:conclusions} presents the conclusions.

\section{Observations}\label{sec:observations}

We observed the YSES 1 system using MagAO-X on the 6.5 metre Magellan Clay telescope at Las Campanas Observatory.
The MagAO-X extreme adaptive optics instrument is designed for high-contrast imaging in the visible to near-infrared \citep{males_magao-x_2024, close_optical_2018}.
It uses a pyramid wavefront sensor in combination with a woofer-tweeter deformable mirror (DM) setup for high-order wavefront control.
It uses an additional DM for non-common-path correction.
The science arm consists of two electron-multiplying CCDs (EMCCDs), {\tt camsci1} and {\tt camsci2}, which perform simultaneous observations in two different filters\footnote{\url{https://magao-x.org/docs/handbook/observers/filters.html}}.

Table \ref{table:1} summarises the observations performed on 7 March 2023.
We \rev{acquired} the observations in pupil-tracking mode to provide angular diversity, enabling the removal of diffraction from the central star \citep{marois_angular_2006}.
We did not use a coronagraph during the observations.
We obtained data in three filters: $r'$, $i'$, and $z'$.
We measured the $i'$ band with {\tt camsci1}, while the $r'$ and $z'$ band were obtained with {\tt camsci2}.
In addition to the science data, we took a set of dark frames for both cameras.

\begin{table*}
\caption{MagAO-X observations.}             
\label{table:1}      
\centering
\def\arraystretch{1.4}
\setlength{\tabcolsep}{11pt}
\begin{tabular}{c c c c c c c c}     
\hline\hline       
Date & Filter & $\lambda_0$, $w_{\mathrm{eff}}$ ($\mu$m) & Detector & {\tt NEXP$\times$NDIT$\times$DIT} (1x1xs) & Field rotation ($\deg$) & Airmass\\ 
\hline                    
   2023/03/07 & $r'$ & 0.614, 0.109 & {\tt camsci2} & 48$\times$1$\times$30 & 8.03 & 1.31\\  
   2023/03/07 & $i'$ & 0.762, 0.126 & {\tt camsci1} & 83$\times$1$\times$30 & 13.10 & 1.32\\
   2023/03/07 & $z'$ & 0.909, 0.132 & {\tt camsci2} & 32$\times$1$\times$30 & 4.76 & 1.34\\
\hline                  
\end{tabular}
\tablefoot{MagAO-X has two science cameras: {\tt camsci1} and {\tt camsci2}.
We set the EMCCD gain to 1.
The parameters $\lambda_0$ and $w_{\mathrm{eff}}$ represent the central wavelength and effective width of the filters.
{\tt NEXP} is the number of exposures, {\tt NDIT} refers to the number of sub-integrations per exposure, and {\tt DIT} is the detector integration time in seconds for a single sub-integration.
The reported airmass is an average value during the observation.
The average seeing was approximately 0.75 arcseconds during the night.}
\end{table*}

We processed the MagAO-X observations using {\tt PynPoint} version 0.11.0 \citep{stolker_pynpoint_2019}.
The {\tt PynPoint} package reduces high-contrast imaging datasets and uses principal component analysis (PCA) to subtract the point spread function (PSF) of the host star \citep{amara_span_2012}.
We began the data reduction by subtracting the median dark frame.
We then rescaled the images along the x axis to account for the different plate scales of the x and y axes, following Table 4 in \citep{long_astrometric_2025}.
We aligned  and centred individual frames on the host star in a series of steps. First, we determined the stellar centre in each frame to pixel accuracy by locating the maximum value after smoothing it with a two-dimensional Gaussian kernel with full width at half maximum (FHWM) of $30 \,\mathrm{mas}$.
Second, we registered the frames by cross-correlating each frame with a subset of randomly selected reference frames to determine the average offset and then resampled the frames using spline interpolation.
Finally, we applied an additional shift to place the star in the centre of the frames.
We removed frames with poor seeing conditions based on aperture photometry of the unsaturated host star PSF to produce the final dataset for \rev{angular differential imaging (ADI)} processing.
We discarded frames in which the host flux was lower than 3-sigma below the maximum measured flux. This led to \rev{6, 17 and 4} frames being removed in the $r'$, $i'$ and $z'$ bands, respectively.

We applied PCA to estimate the PSF of the host star in every frame.
We subtracted the first PCA component and derotated the frames, taking into account rotation corrections of 1.53 and 1.50 degrees for \texttt{camsci1} and \texttt{camsci2}, respectively, to orient north upwards \citep{long_astrometric_2025}.
We took the median values to produce the resulting images.

\begin{figure*}
    \centering
    \includegraphics[trim={2.1cm 1cm 2.5cm 2cm}, clip, width=\linewidth]{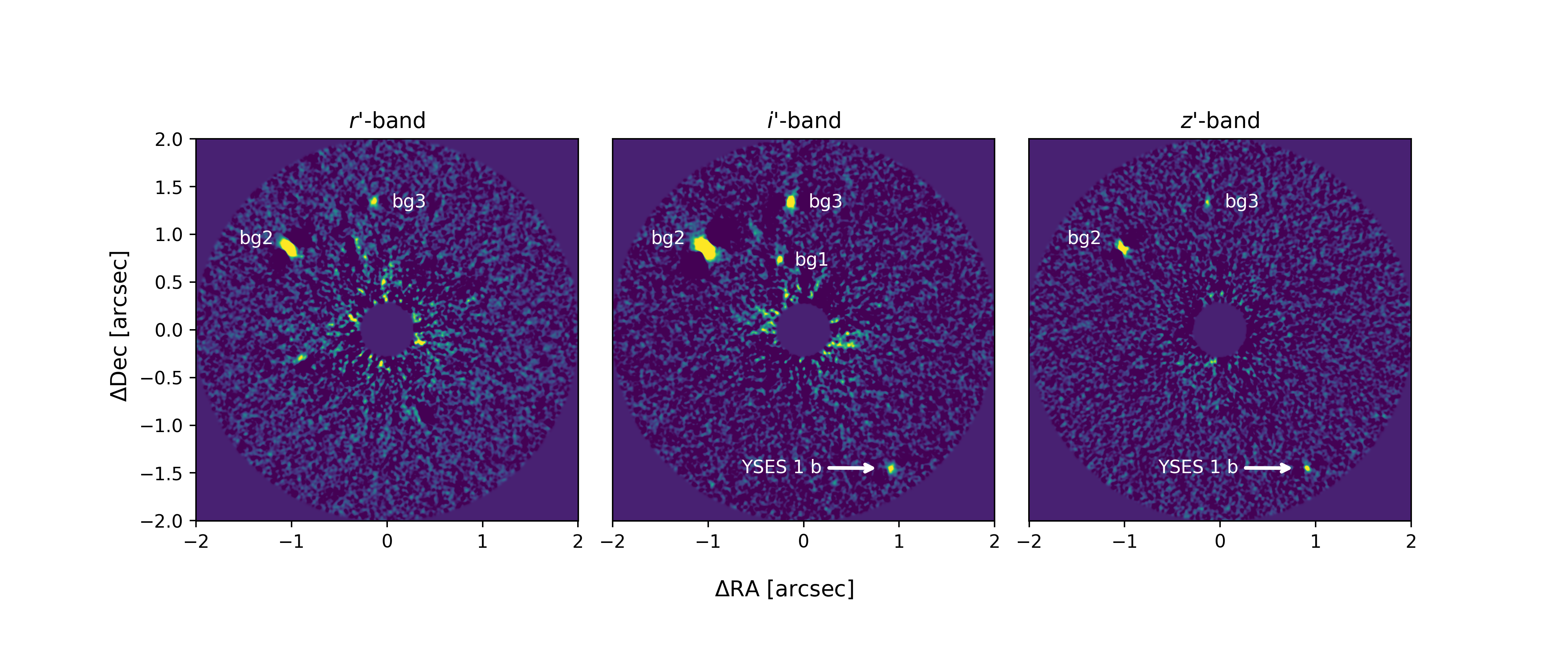}
    \caption{Reduced MagAO-X images of the YSES 1 system in the $r'$, $i'$ and $z'$ filters after PCA-based PSF subtraction and Gaussian filtering.
    The point sources on the top are confirmed background stars \citep{bohn_young_2020}.
    YSES 1 b is recovered in the $i'$ and $z'$ bands with an S/N of 13.0 and 11.2, respectively.}
    \label{fig:magaox}
\end{figure*}

Figure \ref{fig:magaox} shows the final images for each filter.
We applied an additional Gaussian smoothing with a $30 \,\mathrm{mas}$ FWHM kernel to reduce the high-frequency noise left over from the PSF subtraction and to highlight the companion.
We clearly detect YSES 1 b in the $i'$ and $z'$ filters with signal-to-noise ratio (S/N) of $13.0$ and $11.2$, respectively. We estimated the S/N by comparing the companion signal to the noise in an annulus at the same separation, which at these distances is assumed to have a normal distribution.
We do not detect YSES 1 b in the $r'$ filter. 

We obtained the photometry and astrometry of YSES 1 b in the $i'$ and $z'$ bands using the negative artificial planet injection technique implemented in {\tt PynPoint}, in which we subtracted a template PSF at the approximate location of the companion in each frame to remove its signal in the post-processed image \citep{stolker_miracles_2020}.
We then determined the contrast and position of the companion by scaling and moving the PSF template to minimise the residuals.

We estimated the companion PSF in each frame based on the host star PSF by selecting a region of $0.35 \,\mathrm{arcsec}$ around the star.
We then performed a simplex minimisation to determine a first estimate of the best-fit contrast and position.
We used these values as initial conditions for a Markov chain Monte Carlo (MCMC) analysis to retrieve the posterior distributions of the parameters \citep{mackay_information_nodate}.
We selected a total of 50 walkers and 2000 steps per walker for the sampling.
We applied a burn-in of 120 samples and thinned the chain by selecting  every 18th sample based on autocorrelation analysis.
We show the posterior distributions in Appendix \ref{mcmc-posterior}.
In addition to the MCMC estimates, we also analysed possible systematic errors using the injection of artificial planets as described in \citet{stolker_miracles_2020}.
 Appendix \ref{offset} details the distributions for the contrast, separation, and position angle offsets.
We produced the final contrast, separation, and position angle by adding the bias estimates and adding the uncertainties in quadrature.
Table \ref{table:2} provides a summary of the results for both the $i'$ and $z'$ filters.
The separation and position angle are consistent with the GRAVITY measurements reported in \citet{roberts_new_2025}.

Since there is no detection in the $r'$ filter, we approximated the flux as a Gaussian with zero mean and a standard deviation equal to the $1\sigma$ upper limit. Using artificial planet injection at the separation of YSES 1 b to estimate the amount of self-subtraction, we derive a $1\sigma$ magnitude-contrast upper limit of $12.24 \,\mathrm{mag}$.

\begin{table*}
\caption{YSES 1 b contrast and astrometry estimates from MagAO-X images.}             
\label{table:2}      
\centering      
\def\arraystretch{1.4}
\setlength{\tabcolsep}{8pt}
\begin{tabular}{c c c c c c}
\hline\hline       
                      
Filter & Measured contrast ($\mathrm{mag}$) & Bias offset ($\mathrm{mag}$)& Final contrast ($\mathrm{mag}$)& Separation ($\mathrm{mas}$) & Position angle ($\deg$) \\ 
\hline                    
   $r'$ & $12.24$ (1$\sigma$ upper limit) &  &  & & \\  
   $i'$ & $10.09 \pm 0.07$ & $-0.07\pm0.07$ & $10.02\pm0.10$ & $1713 \pm 3.6$ & $212.20 \pm 0.10$ \\
   $z'$ & $8.63 \pm 0.08$ & $-0.02\pm0.08$ & $8.61\pm0.11$ & $1712 \pm 2.8$ & $212.26 \pm 0.06$\\
\hline                  
\end{tabular}
\tablefoot{The error bars on the astrometry are likely underestimated.
Propagation of the plate-scale uncertainties from \citep{long_astrometric_2025} increases the error bars on the separation to 7 mas, but this is an underestimate because we observe relatively large variations between observing runs.}
\end{table*}

\section{Analysis}\label{sec:analysis}

We revisited the physical modelling of YSES 1 b using a combination of the MagAO-X photometry obtained in Section \ref{sec:observations} and archival SPHERE and NACO data \citep{bohn_young_2020}. 
Here, we present the SED fitting and mass estimation.

\subsection{Photometric calibration}

\begin{figure} 
    \centering
    \includegraphics[width=\linewidth]{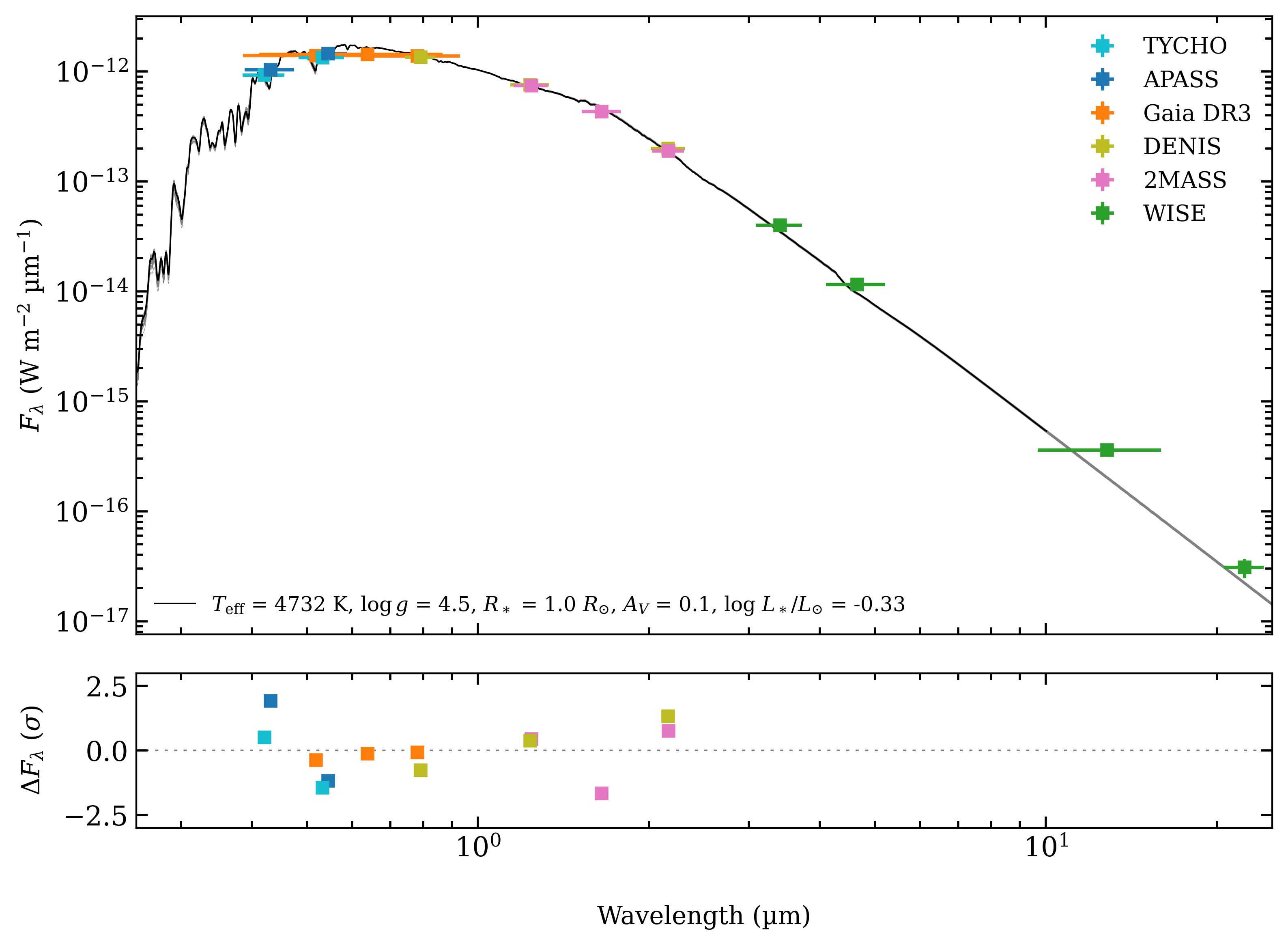}
    \caption{YSES 1 SED fit.
    The photometric points are colour-coded according to their data source.
    The black curve shows the best-fit BT-Settl-CIFIST model with a spectral resolution of 100 $(\chi^2_{\nu}=1.28)$.
    We flagged the WISE photometric points as showing an excess and excluded them from the stellar fit.}
    \label{fig:yses1}
\end{figure}

Converting the derived contrasts in the MagAO-X filters to apparent magnitudes requires a flux calibration of the host star.
To achieve this we performed an SED fit to existing photometric data.
For the stellar atmosphere we employed the BT-Settl-CIFIST model \citep{allard_bt-settl_2013}.
We used the {\tt species} package to estimate the posterior distributions of the model parameters using a nested-sampling approach \citep{stolker_miracles_2020}.
Figure \ref{fig:yses1} shows the results.
We used 17 photometric points from TYCHO \citep{Hog2000}, APASS \citep{APASSDR9}, GAIA \citep{Gaia_DR3}, DENIS \citep{Epchtein1999}, 2MASS \citep{Cutri2003} and WISE \citep{Cutri2012}. 
We set the distance of the system to $94.067\pm0.105 \,(\mathrm{pc})$ \citep{bailer-jones_estimating_2021}.
An initial \rev{Virtual Observatory SED Analyzer (VOSA)} fit detected infrared excess from $W1$ onwards; we therefore discarded the WISE data for the final SED fit \citep{bayo_vosa_2008}.
The stellar atmospheric parameters included in the fit are the effective temperature ($T_{\mathrm{eff}}$), surface gravity ($\log g$), stellar radius ($\mathrm{R_*}$), and extinction in the visible band ($A_{\mathrm{V}}$).
Table {\ref{table:5}} summarises the derived values from the fit and Appendix \ref{stellar-sed-posterior} presents the corresponding posterior distributions.
The reduced chi-square of the best SED fit is 1.28.
We took this best-fit model to calculate the synthetic photometry in the MagAO-X filters and find apparent magnitudes of 10.50, 9.93, and 9.64 for the $r'$, $i'$, and $z'$ filters, respectively.
We use these values throughout the remainder of this work to convert MagAO-X contrast into flux.

\begin{table}[ht]
\caption{Stellar parameters derived for YSES 1 from an SED fit.}             
\label{table:5}      
\centering          
\def\arraystretch{1.4}
\setlength{\tabcolsep}{14pt}
\begin{tabular}{c c c}
\hline
Parameter & Prior (uniform) & Value \\
\hline
$T_{\mathrm{eff}}$ ($\mathrm{K}$) & $[4000 - 6000]$ & $4708^{+30}_{-33}$\\ 
$\log g$ ($\mathrm{dex}$)& $[2.5 - 5.5]$ & $4.46^{+0.38}_{-0.46}$\\
$R_{\mathrm{*}}$ ($\mathrm{R_{\hbox{$\odot$}}}$) & $[0.5 - 1.5]$ & $1.02^{+0.01}_{-0.01}$\\
$A_{\mathrm{V}}$ ($\mathrm{mag}$) & $[0 - 0.102]$ & $0.06^{+0.03}_{-0.04}$\\
\hline
$\log L_{\mathrm{*}}/L_{\hbox{$\odot$}}$ ($\mathrm{dex}$) & & $-0.34^{+0.01}_{-0.01}$\\
\hline
$\chi^2_{\nu}$ & & 1.28\\
\hline                  
\end{tabular}
\tablefoot{The prior on the extinction is the 95\% range estimated for YSES 2 based on the STILISM 3D reddening maps \citep{bohn_discovery_2021}.}
\end{table}

\subsection{SED fit} \label{section:spectral}

In addition to MagAO-X data, we used SPHERE and NACO photometry to perform the SED fits.
For the SPHERE data we used the measurements in the $Y2, Y3, J2, J3, H2, H3, K1 $, and $K2$ filters \citep{beuzit_sphere_2019, bohn_young_2020}.
%
We discarded the $J$, $H$, and $K_s$ measurements due to unstable atmospheric conditions during the observations.
We included both $L'$ and $M'$ data from the NACO instrument \citep{rousset_naos_2003, lenzen_naos-conica_2003, bohn_young_2020}.
We used a forward-modelling approach to fit a grid of theoretical atmospheric spectra to the data.
This was carried out using the {\tt species} package, which employs a \rev{nested-sampling} approach to estimate the posterior distributions of the model parameters \citep{stolker_miracles_2020}.
For the model grid we used BT-Settl-CIFIST.
We fitted the effective temperature ($T_{\mathrm{eff}}$), surface gravity ($\mathrm{\log g}$), and radius ($R_{\mathrm{p}}$).
We also report the resulting bolometric luminosity ($\log L_{\mathrm{p}}/L_{\hbox{$\odot$}}$).

In addition to an atmospheric model, we also considered a model to describe the CPD.
Our CPD model includes dust extinction and a single blackbody component.
We used the G23 dust extinction model with an average Milky Way value of 3.1 for the total-to-selective extinction parameter $R_\mathrm{V}$ \citep{gordon_one_2023}.
We assumed that this interstellar medium (ISM)  model provides a sufficient initial approximation for the CPD dust, since any extinction by the CPD would occur in the upper disc layer, where dust grains are small.
The only fitted parameter is the extinction in the optical band, $A_V$.
We considered other extinction models, such as a power law or log-normal dust distributions, but they did not significantly change the conclusions.
The blackbody component is described by an effective blackbody temperature ($T_{\mathrm{disk\_bb}}$) and a blackbody radius ($R_{\mathrm{disk\_bb}}$).
In addition to the disc luminosity ($\log L_{\mathrm{disk}}/L_{\hbox{$\odot$}}$), we also derived $\log R_{ring}/\mathrm{R_J}$, the radius that a narrow ring of blackbody particles with temperature $T_{\mathrm{disk\_bb}}$ would have around an object with luminosity $\log L_{\mathrm{p}}/L_{\hbox{$\odot$}}$ \citep{kennedy_two-temperature_2014}.

Table \ref{table:3} shows the parameter values derived from the posterior distributions, with and without the addition of the CPD model. 
We present the posterior distributions themselves in Appendix \ref{sed-posterior}. 
Figures \ref{fig:no-cpd} and \ref{fig:cpd} show the respective SED fits, with the maximum likelihood parameters shown in the plots.
Including the CPD model significantly improves the fit to the photometric data.
The reduced chi-square value of the best-fit decreases from 3.75 to 1.46.
Furthermore, the evidence, $\log (Z)$, increases from 448 to 460, which indicates a strong preference for the model including CPD dust extinction and a blackbody.
This substantially changes the derived physical parameters, most notably the effective temperature and radius.
The effective temperature increases from a median value of $1784 \,\mathrm{K}$ to $2854 \,\mathrm{K}$.
The estimated radius decreases from 2.45 $\mathrm{R_J}$ to 1.58 $\mathrm{R_J}$.

\begin{table*}
\caption{Parameters derived for YSES 1 b from forward modelling with and without a CPD.}             
\label{table:3}      
\centering          
\def\arraystretch{1.4}
\setlength{\tabcolsep}{27pt}
\begin{tabular}{c c c c}
\hline
Parameter & Prior (uniform) & BT-Settl-CIFIST & BT-Settl-CIFIST + CPD \\
\hline
$T_{\mathrm{eff}}$ ($\mathrm{K}$) & $[1500 - 4000]$ & 1784$^{+17}_{-20}$ & 2854$^{+110}_{-94}$\\ 
$\log g$ ($\mathrm{dex}$)& $[2.5 - 5.5]$ & 5.23$^{+0.06}_{-0.06}$ & 4.31$^{+0.42}_{-0.43}$\\
$R_{\mathrm{p}}$ ($\mathrm{R_J}$) & $[1 - 5]$ & 2.45$^{+0.07}_{-0.06}$ & 1.58$^{+0.06}_{-0.07}$\\
$A_V$ (mag)& $[1 - 10]$ & - & 5.17$^{+0.32}_{-0.32}$\\
$T_{\mathrm{disk\_bb}}$ ($\mathrm{K}$) & $[30 - 500]$ & - & 339$^{+90}_{-147}$\\
$R_{\mathrm{disk\_bb}}$ ($\mathrm{R_J}$) & $[1 - 100]$ & - & 41.85$^{+36}_{-23}$ \\
\hline
$\log L_{\mathrm{p}}/L_{\hbox{$\odot$}}$ ($\mathrm{dex}$) & & -3.24$^{+0.01}_{-0.01}$ & -2.80$^{+0.04}_{-0.04}$ \\
$\log L_{\mathrm{disk}}/L_{\hbox{$\odot$}}$ ($\mathrm{dex}$) & & - & -3.69$^{+0.38}_{-1.13}$ \\
$\log R_{\mathrm{ring}}/\mathrm{R_J}$ ($\mathrm{dex}$) & & - & $1.75^{+0.50}_{-0.20}$\\
\hline
$\log (Z)$ & & 448.22$\pm0.07$ & 459.84$\pm0.07$ \\
$\chi^2_{\nu}$ & & 3.75 & 1.46\\
\hline                  
\end{tabular}
\end{table*}

\begin{figure*} 
    \centering
    \includegraphics[width=\linewidth]{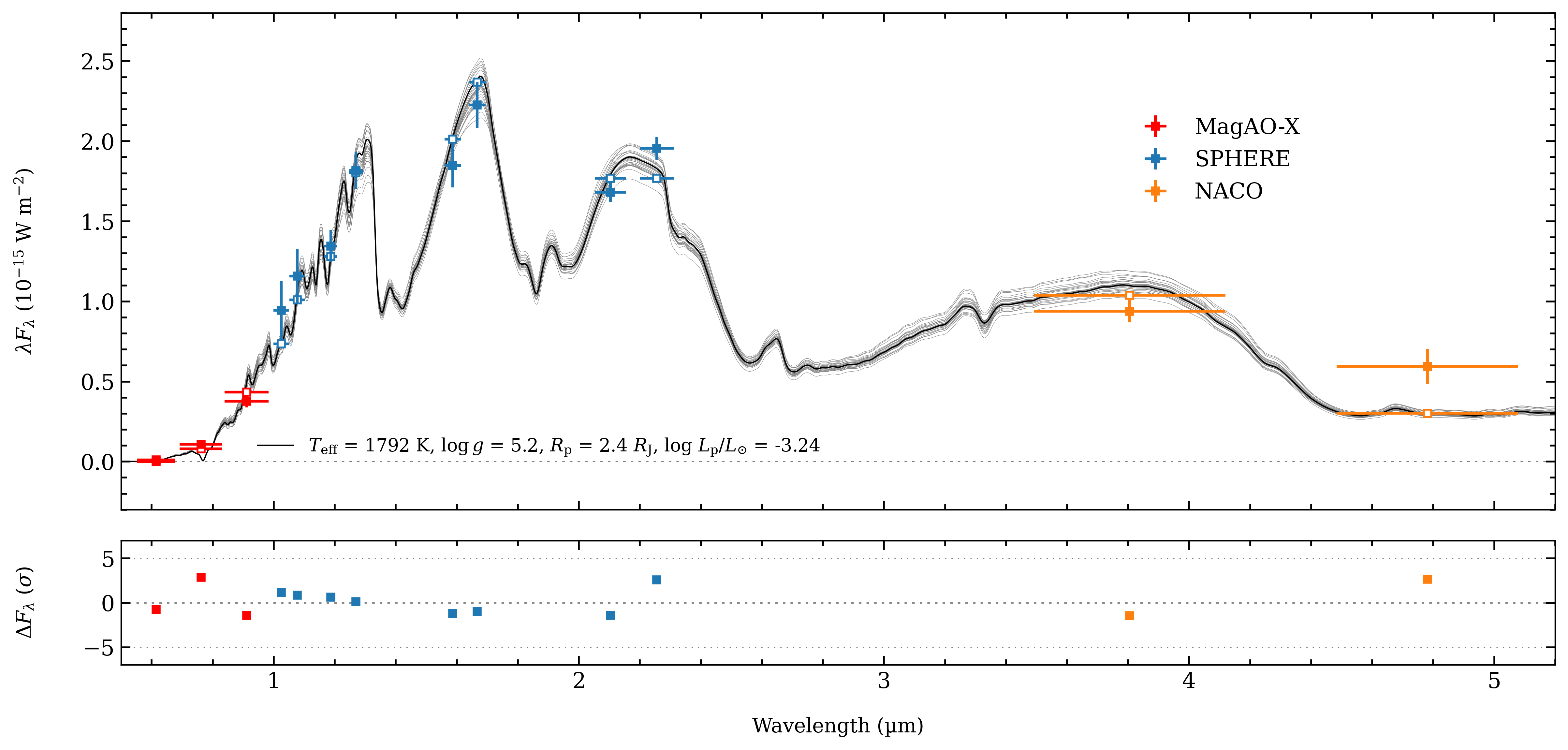}
    \caption{Top: Spectral energy distribution (SED) fit of YSES 1 b using MagAO-X, SPHERE, and NACO data using only an atmospheric model.
    The black line shows the best-fit model $(\chi^2_{\nu}=3.37)$.
    The grey lines show 30 random samples drawn from the posterior distribution.
    The filled markers show the original data points, while the white markers show the synthetic photometry derived from the best-fit model.
    Bottom: Deviation between data and synthetic photometry in units of the standard deviation.}
    \label{fig:no-cpd}
\end{figure*}

\begin{figure*} 
    \centering
    \includegraphics[width=\linewidth]{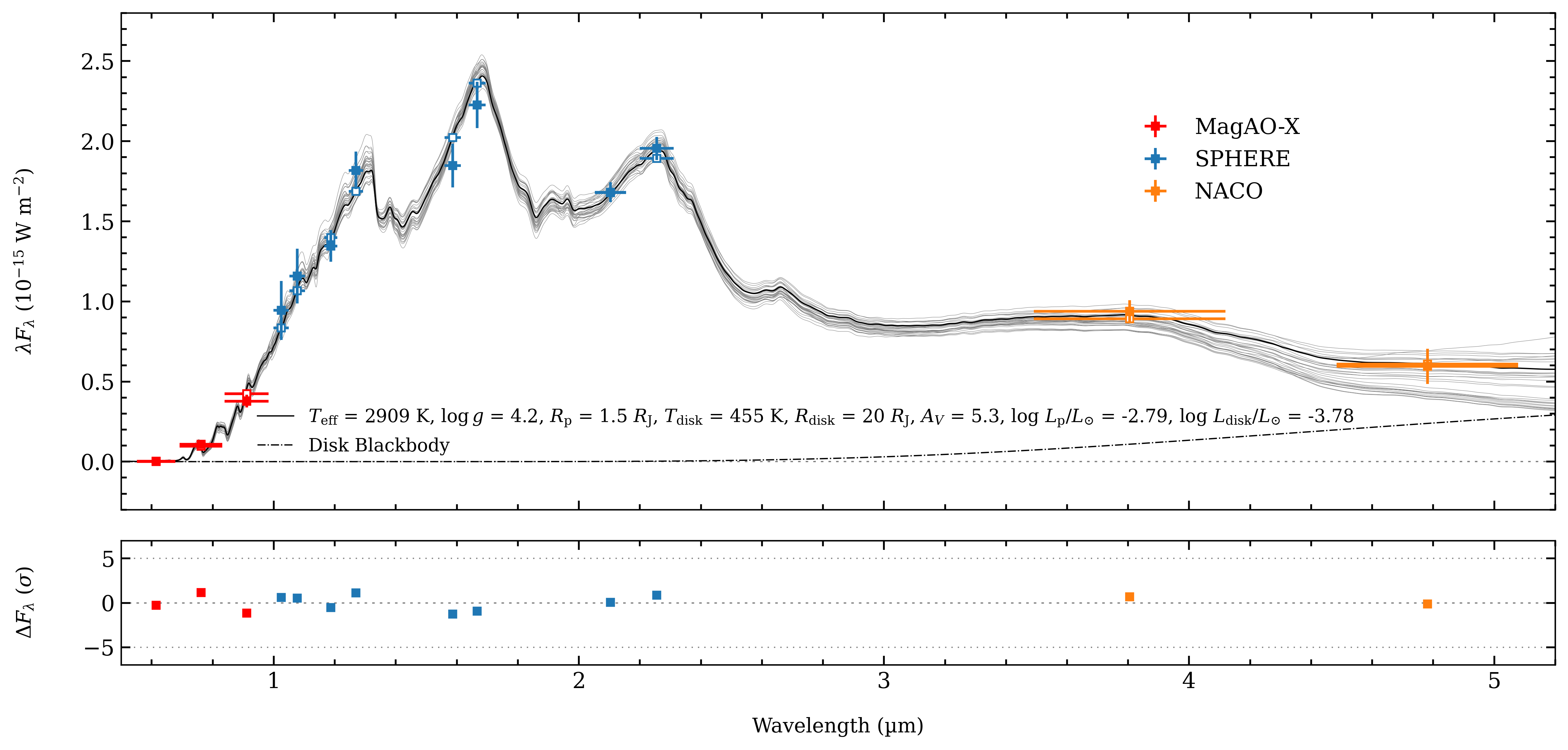}
    \caption{Top: Spectral energy distribution (SED) fit of YSES 1 b using MagAO-X, SPHERE, and NACO data using an atmospheric model in combination with a dust extinction model and blackbody radiation to represent the CPD.
    The black line shows the best-fit $(\chi^2_{\nu}=1.46)$.
    The dash-dotted line shows the contribution of the blackbody component.
    The grey lines show 30 random samples drawn from the posterior distribution.
    The filled markers are the original data points, while the white markers show the synthetic photometry derived from the best-fit model.
    Bottom: Deviation between data and synthetic photometry in units of the standard deviation.}
    \label{fig:cpd}
\end{figure*}

\subsection{Mass estimate}

Using the median bolometric luminosity ($\log L_{\mathrm{p}}/L_{\hbox{$\odot$}}$) from Table \ref{table:3} and the age of the system, we obtain an estimate of the mass of the object by fitting an evolutionary model.
For the system age we considered two values: the original $16.7\pm1.4 \,\mathrm{Myr}$ \citep{bohn_young_2020} and a more recent estimate of $27\pm3 \,\mathrm{Myr}$ \citep{wood_tess_2023}.
We used the BT-Settl isochrones as our evolutionary model \citep{allard_model_2011}.
Using the nested-sampling fitting procedure of \texttt{species}, we derived posterior distributions for the following parameters: age, mass ($M$), effective temperature ($T_{\mathrm{eff}}$), radius ($R_{\mathrm{p}}$) and surface gravity ($\mathrm{\log g}$).
Table \ref{table:4} summarises the results and the full posterior distributions can be found in Appendix \ref{evolutionary-posterior}.
The median estimated mass is either $25.7 \,\mathrm{M_J}$ or $41.6 \,\mathrm{M_J}$, depending on the age of the system.
 Although evolutionary modelling fits a single luminosity parameter and therefore introduces an ambiguity in the derived effective temperature and radius, the results can be compared with Table \ref{table:3}.
The values corresponding to the age of $27 \,\mathrm{Myr}$ are in better agreement with those derived from the SED fit, indicating that the older age is more consistent with these results.

\begin{table}
\caption{Evolutionary fit for the estimated ages of 16.7 Myr \citep{bohn_young_2020} and 27 Myr  \citep{wood_tess_2023}.}             
\label{table:4}      
\centering          
\def\arraystretch{1.4}
\setlength{\tabcolsep}{19pt}
\begin{tabular}{c c c}
\hline
Age & 16.7$\pm1.4$ Myr & 27$\pm3$ Myr \\
\hline
$M$ ($\mathrm{M_J}$)& 25.7$^{+4.1}_{-3.6}$ & 41.6$^{+3.6}_{-3.4}$\\
$T_{\mathrm{eff}}$ ($\mathrm{K}$) & 2512$^{+49}_{-46}$ & 2638$^{+38}_{-43}$\\ 
$R_{\mathrm{p}}$ ($\mathrm{R_J}$) & 2.08$^{+0.06}_{-0.06}$ & 1.89$^{+0.06}_{-0.05}$\\
log $g$ ($\mathrm{dex}$) & 4.17$^{+0.08}_{-0.08}$ & 4.47$^{+0.04}_{-0.05}$\\
\hline                  
\end{tabular}
\end{table}

\section{Discussion} \label{sec:discussion}

Using only the atmospheric model in Table \ref{table:3}, we  compared the retrieved parameters with those in Figure 5 of \citet{bohn_young_2020} to test \rev{if our SED fit with MagAO-X data is consistent}. 
The effective temperatures are in good agreement (1784 versus 1727 $\mathrm{K}$). 
We find median $\mathrm{\log g}$ and radius values of 5.23 $\mathrm{dex}$ and 2.45 $ \mathrm{R_J}$, which deviate from the values of 3.91 $\mathrm{dex}$ and 3.0 $\mathrm{R_J}$ reported by \citet{bohn_young_2020}, although both remain consistent within the uncertainties.
However, \citet{bohn_young_2020} achieve a better fit, especially at wavelengths longer than $2\,\mu\mathrm{m}$.
This is due to their use of the original BT-Settl model rather than the CIFIST release. 
We adopted the CIFIST release because it uses updated solar abundances \citep{caffau_solar_2011}.

Regardless of the atmospheric model used, one key issue is that the estimated radius is still considerably larger than that expected for a low-mass companion \citep{baraffe_new_2015}.
The inclusion of a blackbody component by \citet{hoch_silicate_2025} to model the CPD still results in a radius of 3 $\mathrm{R_J}$ from forward modelling.
Using the CPD model in this work, which includes both dust extinction and a blackbody component, reduces the radius to 1.58 $\mathrm{R_J}$, in much better agreement with the expected radius of a substellar companion of this age.
The CPD cavity-size estimate $R_{\mathrm{ring}}$ corresponding to 56$^{+178}_{-36} \,\mathrm{R_J}$ is on the same order of magnitude as the values of $40\pm0.7\,\mathrm{R_J}$ and $33\pm3\,\mathrm{R_J}$ reported for GQ Lup b and Delorme 1 (AB) b \citep{cugno_mid-infrared_2024, malin_jwst-tst_2025}. Nevertheless, the value derived for YSES 1 b from the SED is much less constrained and therefore difficult to compare.
The estimated masses of $25.7 \,\mathrm{M_J}$ or $41.6 \,\mathrm{M_J}$ are significantly higher than the $14 \,\mathrm{M_J}$ reported in \citet{bohn_young_2020}.
Using the 13 $\mathrm{M_J}$ threshold, although it does not fully capture the separation between brown dwarfs and planets \citep{deeg_definition_2018}, the newly derived mass more strongly supports a brown dwarf interpretation for YSES 1 b.

\begin{figure} 
    \centering
    \includegraphics[width=\linewidth]{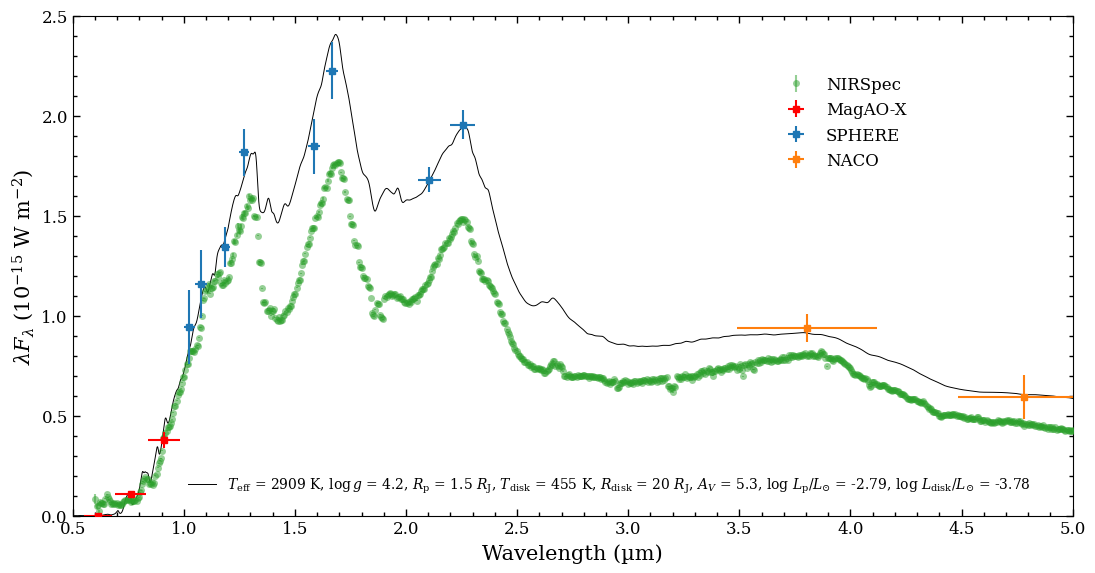}
    \caption{Comparison of ground-based photometry and our best-fit model to the JWST NIRSpec data.}
    \label{fig:comparison}
\end{figure}

A comparison between the JWST \citep{hoch_silicate_2025} flux and the existing SPHERE and NACO photometry shows a significant difference.
This is shown in Figure \ref{fig:comparison}, where we plot the JWST/NIRSpec data and the ground-based photometry.
The ground-based and space-based data are consistent up to 1.2 $\mu$m.
At wavelengths longer than 1.2 $\mu\mathrm{m}$, JWST shows a significantly lower flux than the ground-based measurements.
The difference between the best-fit model and the JWST NIRSpec spectrum in this wavelength range is approximately 15-45\%.
We re-reduced the SPHERE data to search for an explanation for this discrepancy, but we found the same contrast levels as before.
In addition, we compared the contrast in the SPHERE $K1$ and $K2$ filters to recently published GRAVITY $\mathrm{K}$-band spectra and found them to be consistent \citep{kammerer_exogravity_2025}.
Variability in YSES 1 b may be considered in this context. 
Other young planet and brown dwarf companions show variability due to cloud cover. 
Nevertheless, the reported levels of variability are generally lower, ranging from the percent level 
to the 26\% percent reported by \cite{radigan_large-amplitude_2012} \citep{biller_time_2017, sutlieff_measuring_2023}.
\rev{Furthermore, }the newly derived effective temperature is too high \rev{to expect} cloud condensation to occur, so the variability would need to arise from a different source.
Variability in the dust extinction would lead to the largest discrepancies at short wavelengths, but the JWST data are consistent at wavelengths below 1.2 $\mu\mathrm{m}$.
In addition, the ground-based data were taken at different epochs and show consistent results, which makes a strong variability scenario less favourable.
Therefore, future work and observations at different epochs are needed to validate the cause of the discrepancy.

\section{Conclusions} \label{sec:conclusions}

We obtained data on YSES 1 b in the $r'$, $i'$ and $z'$ bands using MagAO-X.
We clearly detect YSES 1 b in the $i'$ and $z'$ filters, while we derive only an upper limit in the $r'$ band.
The MagAO-X data are consistent with models fitted to the existing SPHERE and NACO photometry.
We included a model of the CPD which was recently confirmed using JWST \citep{hoch_silicate_2025}.
Adding dust extinction and blackbody emission led to an SED that, compared to the initial characterisation of YSES 1 b by \citet{bohn_young_2020}, is significantly more massive (25.7$^{+4.1}_{-3.6}$ or 41.6$^{+3.6}_{-3.4}$ $\mathrm{M_J}$ versus 14$\pm 3$ $\mathrm{M_J}$), warmer (2854$^{+110}_{-94}$ $\mathrm{K}$ versus 1727$^{+172}_{-127}$ $\mathrm{K}$), and smaller (1.58$^{+0.06}_{-0.07}$ $\mathrm{R_J}$ versus 3.0$^{+0.2}_{-0.7}$ $\mathrm{R_J}$).
This suggests that YSES 1 b is more likely to lie in the brown dwarf regime rather than the planet regime.
It also resolves the previously identified large-radius anomaly, as the newly determined radius is consistent with those of other young planets and brown dwarfs.

\begin{acknowledgements}
This work is financially supported by the Netherlands Organisation for Scientific Research NWO (Vidi 213.149). 
We thank Kielan Hoch for sharing the JWST/NIRSpec spectrum.
We are very grateful for support from the NSF MRI Award \#1625441. 
The Phase II upgrade program is made possible by the generous support of the Heising-Simons Foundation.
The development of CACAO is supported by NSF Award \#2410616.
\end{acknowledgements}

%
%


\bibliographystyle{aa} 
\bibliography{references}

@article{stolker_pynpoint_2019,
	title = {{PynPoint}: a modular pipeline architecture for processing and analysis of high-contrast imaging data},
	volume = {621},
	copyright = {https://www.edpsciences.org/en/authors/copyright-and-licensing},
	issn = {0004-6361, 1432-0746},
	shorttitle = {{PynPoint}},
	url = {https://www.aanda.org/10.1051/0004-6361/201834136},
	doi = {10.1051/0004-6361/201834136},
	language = {en},
	urldate = {2025-08-18},
	journal = {A\&A},
	author = {Stolker, T. and Bonse, M. J. and Quanz, S. P. and Amara, A. and Cugno, G. and Bohn, A. J. and Boehle, A.},
	month = jan,
	year = {2019},
	pages = {A59},
}

@article{malin_jwst-tst_2025,
	title = {{JWST}-{TST} {High} {Contrast}: {Medium}-resolution spectroscopy reveals a carbon-rich circumplanetary disk around the young accreting exoplanet {Delorme} 1 {AB} b},
	shorttitle = {{JWST}-{TST} {High} {Contrast}},
	url = {http://arxiv.org/abs/2510.07253},
	doi = {10.48550/arXiv.2510.07253},
	language = {en},
	urldate = {2026-03-23},
	author = {Mâlin, Mathilde and Ward-Duong, Kimberly and Grant, Sierra L. and Arulanantham, Nicole and Tabone, Benoit and Pueyo, Laurent and Perrin, Marshall and Balmer, William O. and Betti, Sarah and Chen, Christine H. and Debes, John H. and Girard, Julien H. and Hoch, Kielan K. W. and Kammerer, Jens and Lu, Cicero and Rebollido, Isabel and Rickman, Emily and Robinson, Connor and Worthen, Kadin and Marel, Roeland P. van der and Lewis, Nikole K. and Seager, Sara and Valenti, Jeff A. and Soummer, Remi},
	month = dec,
	year = {2025},
    volume = {704},
    journal = {A\&A},
    pages = {A181},
}

@article{radigan_large-amplitude_2012,
	title = {{LARGE}-{AMPLITUDE} {VARIATIONS} {OF} {AN} {L}/{T} {TRANSITION} {BROWN} {DWARF}: {MULTI}-{WAVELENGTH} {OBSERVATIONS} {OF} {PATCHY}, {HIGH}-{CONTRAST} {CLOUD} {FEATURES}},
	volume = {750},
	issn = {0004-637X, 1538-4357},
	shorttitle = {{LARGE}-{AMPLITUDE} {VARIATIONS} {OF} {AN} {L}/{T} {TRANSITION} {BROWN} {DWARF}},
	url = {https://iopscience.iop.org/article/10.1088/0004-637X/750/2/105},
	doi = {10.1088/0004-637X/750/2/105},
	language = {en},
	number = {2},
	urldate = {2026-03-30},
	journal = {ApJ},
	author = {Radigan, Jacqueline and Jayawardhana, Ray and Lafrenière, David and Artigau, Étienne and Marley, Mark and Saumon, Didier},
	month = may,
	year = {2012},
	pages = {105},
}

@article{cugno_mid-infrared_2024,
	title = {Mid-infrared {Spectrum} of the {Disk} around the {Forming} {Companion} {GQ} {Lup} {B} {Revealed} by {JWST}/{MIRI}},
	volume = {966},
	issn = {2041-8205, 2041-8213},
	url = {https://iopscience.iop.org/article/10.3847/2041-8213/ad3cbc},
	doi = {10.3847/2041-8213/ad3cbc},
	language = {en},
	number = {1},
	urldate = {2026-03-23},
	journal = {ApJL},
	author = {Cugno, Gabriele and Patapis, Polychronis and Banzatti, Andrea and Meyer, Michael and Dannert, Felix A. and Stolker, Tomas and MacDonald, Ryan J. and Pontoppidan, Klaus M.},
	month = may,
	year = {2024},
	pages = {L21},
}

@article{julo_stellar_2025,
	title = {Stellar halo subtraction alternative for accreting companions' characterization with integral field spectroscopy: {Analytical} and on-sky demonstration on the {PDS70}, {HTLup}, and {YSES1} systems},
	volume = {703},
	copyright = {https://creativecommons.org/licenses/by/4.0},
	issn = {0004-6361, 1432-0746},
	shorttitle = {Stellar halo subtraction alternative for accreting companions' characterization with integral field spectroscopy},
	url = {https://www.aanda.org/10.1051/0004-6361/202453231},
	doi = {10.1051/0004-6361/202453231},
	language = {en},
	urldate = {2026-02-26},
	journal = {A\&A},
	author = {Julo, R. and Bonnefoy, M. and Chatelain, F. and Flasseur, O. and Michel, O. J. J. and Jorquera, S. and Delorme, P. and Chauvin, G.},
	month = nov,
	year = {2025},
	pages = {A205},
}

@inproceedings{close_optical_2018,
author = {Laird M. Close and Jared R. Males and Olivier Durney and Corwynn Sauve and Maggie Kautz and Alex Hedglen and Lauren Schatz and Jennifer Lumbres and Kelsey Miller and Kyle Van Gorkom and Madison Jean and Victor Gasho},
title = {{Optical and mechanical design of the extreme AO coronagraphic instrument MagAO-X}},
volume = {10703},
booktitle = {Adaptive Optics Systems VI},
publisher = {SPIE},
pages = {107034Y},
year = {2018},
doi = {10.1117/12.2312280},
URL = {https://doi.org/10.1117/12.2312280}
}

@book{mackay_information_nodate,
	title = {Information {Theory}, {Inference}, and {Learning} {Algorithms}},
	language = {en},
	publisher = {Cambridge University Press},
	author = {MacKay, David J C},
    year = {2003}
}

@article{bailer-jones_estimating_2021,
	title = {Estimating distances from parallaxes. {V}: {Geometric} and photogeometric distances to 1.47 billion stars in {Gaia} {Early} {Data} {Release} 3},
	volume = {161},
	issn = {0004-6256, 1538-3881},
	shorttitle = {Estimating distances from parallaxes. {V}},
	url = {http://arxiv.org/abs/2012.05220},
	doi = {10.3847/1538-3881/abd806},
	language = {en},
	number = {3},
	urldate = {2025-12-12},
	journal = {AJ},
	author = {Bailer-Jones, C. A. L. and Rybizki, J. and Fouesneau, M. and Demleitner, M. and Andrae, R.},
	month = mar,
	year = {2021},
	pages = {147},
}

@article{bohn_discovery_2021,
	title = {Discovery of a directly imaged planet to the young solar analog {YSES} 2},
	volume = {648},
	copyright = {https://www.edpsciences.org/en/authors/copyright-and-licensing},
	issn = {0004-6361, 1432-0746},
	url = {https://www.aanda.org/10.1051/0004-6361/202140508},
	doi = {10.1051/0004-6361/202140508},
	language = {en},
	urldate = {2025-12-09},
	journal = {A\&A},
	author = {Bohn, Alexander J. and Ginski, Christian and Kenworthy, Matthew A. and Mamajek, Eric E. and Pecaut, Mark J. and Mugrauer, Markus and Vogt, Nikolaus and Adam, Christian and Meshkat, Tiffany and Reggiani, Maddalena and Snik, Frans},
	month = apr,
	year = {2021},
	pages = {A73},
}

@article{Hog2000,
	Adsurl = {http://adsabs.harvard.edu/abs/2000A%26A...363..385H},
	Author = {{H{\o}g}, E. and {Fabricius}, C. and {Makarov}, V.~V. and {Urban}, S. and {Corbin}, T. and {Wycoff}, G. and {Bastian}, U. and {Schwekendiek}, P. and {Wicenec}, A.},
	Journal = {\aap},
	Month = nov,
	Pages = {385},
	Title = {{(Erratum) Letter to the Editor - The Tycho-2 catalogue of the 2.5 million brightest stars}},
	Volume = 363,
	Year = 2000}

@ARTICLE{Gaia_DR3,
       author = {{Gaia Collaboration} and {Vallenari}, A. and {Brown}, A.~G.~A. and {Prusti}, T. and {de Bruijne}, J.~H.~J. and {Arenou}, F. and {Babusiaux}, C. and {Biermann}, M. and {Creevey}, O.~L. and {Ducourant}, C. and {Evans}, D.~W. and {Eyer}, L. and {Guerra}, R. and {Hutton}, A. and {Jordi}, C. and {Klioner}, S.~A. and {Lammers}, U.~L. and {Lindegren}, L. and {Luri}, X. and {Mignard}, F. and {Panem}, C. and {Pourbaix}, D. and {Randich}, S. and {Sartoretti}, P. and {Soubiran}, C. and {Tanga}, P. and {Walton}, N.~A. and {Bailer-Jones}, C.~A.~L. and {Bastian}, U. and {Drimmel}, R. and {Jansen}, F. and {Katz}, D. and {Lattanzi}, M.~G. and {van Leeuwen}, F. and {Bakker}, J. and {Cacciari}, C. and {Casta{\~n}eda}, J. and {De Angeli}, F. and {Fabricius}, C. and {Fouesneau}, M. and {Fr{\'e}mat}, Y. and {Galluccio}, L. and {Guerrier}, A. and {Heiter}, U. and {Masana}, E. and {Messineo}, R. and {Mowlavi}, N. and {Nicolas}, C. and {Nienartowicz}, K. and {Pailler}, F. and {Panuzzo}, P. and {Riclet}, F. and {Roux}, W. and {Seabroke}, G.~M. and {Sordo}, R. and {Th{\'e}venin}, F. and {Gracia-Abril}, G. and {Portell}, J. and {Teyssier}, D. and {Altmann}, M. and {Andrae}, R. and {Audard}, M. and {Bellas-Velidis}, I. and {Benson}, K. and {Berthier}, J. and {Blomme}, R. and {Burgess}, P.~W. and {Busonero}, D. and {Busso}, G. and {C{\'a}novas}, H. and {Carry}, B. and {Cellino}, A. and {Cheek}, N. and {Clementini}, G. and {Damerdji}, Y. and {Davidson}, M. and {de Teodoro}, P. and {Nu{\~n}ez Campos}, M. and {Delchambre}, L. and {Dell'Oro}, A. and {Esquej}, P. and {Fern{\'a}ndez-Hern{\'a}ndez}, J. and {Fraile}, E. and {Garabato}, D. and {Garc{\'\i}a-Lario}, P. and {Gosset}, E. and {Haigron}, R. and {Halbwachs}, J. -L. and {Hambly}, N.~C. and {Harrison}, D.~L. and {Hern{\'a}ndez}, J. and {Hestroffer}, D. and {Hodgkin}, S.~T. and {Holl}, B. and {Jan{\ss}en}, K. and {Jevardat de Fombelle}, G. and {Jordan}, S. and {Krone-Martins}, A. and {Lanzafame}, A.~C. and {L{\"o}ffler}, W. and {Marchal}, O. and {Marrese}, P.~M. and {Moitinho}, A. and {Muinonen}, K. and {Osborne}, P. and {Pancino}, E. and {Pauwels}, T. and {Recio-Blanco}, A. and {Reyl{\'e}}, C. and {Riello}, M. and {Rimoldini}, L. and {Roegiers}, T. and {Rybizki}, J. and {Sarro}, L.~M. and {Siopis}, C. and {Smith}, M. and {Sozzetti}, A. and {Utrilla}, E. and {van Leeuwen}, M. and {Abbas}, U. and {{\'A}brah{\'a}m}, P. and {Abreu Aramburu}, A. and {Aerts}, C. and {Aguado}, J.~J. and {Ajaj}, M. and {Aldea-Montero}, F. and {Altavilla}, G. and {{\'A}lvarez}, M.~A. and {Alves}, J. and {Anders}, F. and {Anderson}, R.~I. and {Anglada Varela}, E. and {Antoja}, T. and {Baines}, D. and {Baker}, S.~G. and {Balaguer-N{\'u}{\~n}ez}, L. and {Balbinot}, E. and {Balog}, Z. and {Barache}, C. and {Barbato}, D. and {Barros}, M. and {Barstow}, M.~A. and {Bartolom{\'e}}, S. and {Bassilana}, J. -L. and {Bauchet}, N. and {Becciani}, U. and {Bellazzini}, M. and {Berihuete}, A. and {Bernet}, M. and {Bertone}, S. and {Bianchi}, L. and {Binnenfeld}, A. and {Blanco-Cuaresma}, S. and {Blazere}, A. and {Boch}, T. and {Bombrun}, A. and {Bossini}, D. and {Bouquillon}, S. and {Bragaglia}, A. and {Bramante}, L. and {Breedt}, E. and {Bressan}, A. and {Brouillet}, N. and {Brugaletta}, E. and {Bucciarelli}, B. and {Burlacu}, A. and {Butkevich}, A.~G. and {Buzzi}, R. and {Caffau}, E. and {Cancelliere}, R. and {Cantat-Gaudin}, T. and {Carballo}, R. and {Carlucci}, T. and {Carnerero}, M.~I. and {Carrasco}, J.~M. and {Casamiquela}, L. and {Castellani}, M. and {Castro-Ginard}, A. and {Chaoul}, L. and {Charlot}, P. and {Chemin}, L. and {Chiaramida}, V. and {Chiavassa}, A. and {Chornay}, N. and {Comoretto}, G. and {Contursi}, G. and {Cooper}, W.~J. and {Cornez}, T. and {Cowell}, S. and {Crifo}, F. and {Cropper}, M. and {Crosta}, M. and {Crowley}, C. and {Dafonte}, C. and {Dapergolas}, A. and {David}, M. and {David}, P. and {de Laverny}, P. and {De Luise}, F. and {De March}, R.},
        title = "{Gaia Data Release 3. Summary of the content and survey properties}",
      journal = {\aap},
         year = 2023,
        month = jun,
       volume = {674},
          eid = {A1},
        pages = {A1},
          doi = {10.1051/0004-6361/202243940},
archivePrefix = {arXiv},
       eprint = {2208.00211},
 primaryClass = {astro-ph.GA},
       adsurl = {https://ui.adsabs.harvard.edu/abs/2023A\&A...674A...1G}
}

@ARTICLE{Epchtein1999,
   author = {{Epchtein}, N. and {Deul}, E. and {Derriere}, S. and {Borsenberger}, J. and 
	{Egret}, D. and {Simon}, G. and {Alard}, C. and {Bal{\'a}zs}, L.~G. and 
	{de Batz}, B. and {Cioni}, M.-R. and {Copet}, E. and {Dennefeld}, M. and 
	{Forveille}, T. and {Fouqu{\'e}}, P. and {Garz{\'o}n}, F. and 
	{Habing}, H.~J. and {Holl}, A. and {Hron}, J. and {Kimeswenger}, S. and 
	{Lacombe}, F. and {Le Bertre}, T. and {Loup}, C. and {Mamon}, G.~A. and 
	{Omont}, A. and {Paturel}, G. and {Persi}, P. and {Robin}, A.~C. and 
	{Rouan}, D. and {Tiph{\`e}ne}, D. and {Vauglin}, I. and {Wagner}, S.~J.
	},
    title = "{A preliminary database of DENIS point sources}",
  journal = {\aap},
     year = 1999,
    month = sep,
   volume = 349,
    pages = {236-242},
   adsurl = {http://adsabs.harvard.edu/abs/1999A%26A...349..236E}
}

@misc{APASSDR9,
       author = {{Henden}, A.~A. and {Templeton}, M. and {Terrell}, D. and {Smith}, T.~C. and {Levine}, S. and {Welch}, D.},
        title = "{VizieR Online Data Catalog: AAVSO Photometric All Sky Survey (APASS) DR9}",
         year = 2016,
        month = jan,
          eid = {II/336},
       adsurl = {https://ui.adsabs.harvard.edu/abs/2016yCat.2336....0H}
}

@MISC{Cutri2012,
       author = {{Cutri}, R.~M. and {Wright}, E.~L. and {Conrow}, T. and {Bauer}, J. and {Benford}, D. and {Brandenburg}, H. and {Dailey}, J. and {Eisenhardt}, P.~R.~M. and {Evans}, T. and {Fajardo-Acosta}, S. and {Fowler}, J. and {Gelino}, C. and {Grillmair}, C. and {Harbut}, M. and {Hoffman}, D. and {Jarrett}, T. and {Kirkpatrick}, J.~D. and {Leisawitz}, D. and {Liu}, W. and {Mainzer}, A. and {Marsh}, K. and {Masci}, F. and {McCallon}, H. and {Padgett}, D. and {Ressler}, M.~E. and {Royer}, D. and {Skrutskie}, M.~F. and {Stanford}, S.~A. and {Wyatt}, P.~L. and {Tholen}, D. and {Tsai}, C.~W. and {Wachter}, S. and {Wheelock}, S.~L. and {Yan}, L. and {Alles}, R. and {Beck}, R. and {Grav}, T. and {Masiero}, J. and {McCollum}, B. and {McGehee}, P. and {Papin}, M. and {Wittman}, M.},
        title = "{Explanatory Supplement to the WISE All-Sky Data Release Products}",
         year = 2012,
        month = mar,
        pages = {1},
       adsurl = {https://ui.adsabs.harvard.edu/abs/2012wise.rept....1C}
}

@article{biller_time_2017,
	title = {The time domain for brown dwarfs and directly imaged giant exoplanets: the power of variability monitoring},
	volume = {13},
	issn = {2167-2857, 2167-2865},
	shorttitle = {The time domain for brown dwarfs and directly imaged giant exoplanets},
	url = {https://www.tandfonline.com/doi/full/10.1080/21672857.2017.1303105},
	doi = {10.1080/21672857.2017.1303105},
	language = {en},
	number = {1},
	urldate = {2025-12-07},
	journal = {AR},
	author = {Biller, Beth},
	month = jan,
	year = {2017},
	pages = {1--27},
}

@article{sutlieff_measuring_2023,
	title = {Measuring the variability of directly imaged exoplanets using vector {Apodizing} {Phase} {Plates} combined with ground-based differential spectrophotometry},
	volume = {520},
	issn = {0035-8711, 1365-2966},
	url = {http://arxiv.org/abs/2301.08689},
	doi = {10.1093/mnras/stad249},
	language = {en},
	number = {3},
	urldate = {2025-12-07},
	journal = {MNRAS},
	author = {Sutlieff, Ben J. and Birkby, Jayne L. and Stone, Jordan M. and Doelman, David S. and Kenworthy, Matthew A. and Panwar, Vatsal and Bohn, Alexander J. and Ertel, Steve and Snik, Frans and Woodward, Charles E. and Skemer, Andrew J. and Leisenring, Jarron M. and Strassmeier, Klaus G. and Charbonneau, David},
	month = feb,
	year = {2023},
	pages = {4235--4257},
}

@article{kennedy_two-temperature_2014,
	title = {Do two-temperature debris discs have multiple belts?},
	volume = {444},
	issn = {1365-2966, 0035-8711},
	url = {http://academic.oup.com/mnras/article/444/4/3164/1021189/Do-twotemperature-debris-discs-have-multiple-belts},
	doi = {10.1093/mnras/stu1665},
	language = {en},
	number = {4},
	urldate = {2025-12-07},
	journal = {MNRAS},
	author = {Kennedy, G. M. and Wyatt, M. C.},
	month = nov,
	year = {2014},
	pages = {3164--3182},
}

@inproceedings{lenzen_naos-conica_2003,
	title = {{NAOS}-{CONICA} first on sky results in a variety of observing modes},
	url = {http://proceedings.spiedigitallibrary.org/proceeding.aspx?doi=10.1117/12.460044},
	doi = {10.1117/12.460044},
	language = {en},
	urldate = {2025-11-30},
	author = {Lenzen, Rainer and Hartung, Markus and Brandner, Wolfgang and Finger, Gert and Hubin, Norbert N. and Lacombe, Francois and Lagrange, Anne-Marie and Lehnert, Matthew D. and Moorwood, Alan F. M. and Mouillet, David},
	month = mar,
	year = {2003},
	pages = {944},
    booktitle = {Proc. {SPIE} 4841},
}

@inproceedings{rousset_naos_2003,
	title = {{NAOS}, the first {AO} system of the {VLT}: on-sky performance},
	doi = {https://doi.org/10.1117/12.459332},
	booktitle = {Proc. {SPIE} 4839},
	author = {Rousset, G. and Lacombe, F. and Puget, P. and Hubin, N. and Gendron, E. and Fusco, T. and Arsenault, R and Charton, J. and Feautrier, P. and Gigan, P. and Kern, P. and Lagrange, A.-M. and Madec, Pierre-Yves and Mouillet, D. and Rabaud, D. and Rabou, P. and Stadler, E. and Zins, G.},
	year = {2003},
}

@article{beuzit_sphere_2019,
	title = {{SPHERE}: the exoplanet imager for the {Very} {Large} {Telescope}},
	volume = {631},
	copyright = {http://creativecommons.org/licenses/by/4.0},
	issn = {0004-6361, 1432-0746},
	shorttitle = {{SPHERE}},
	url = {https://www.aanda.org/10.1051/0004-6361/201935251},
	doi = {10.1051/0004-6361/201935251},
	language = {en},
	urldate = {2025-11-25},
	journal = {A\&A},
	author = {Beuzit, J.-L. and Vigan, A. and Mouillet, D. and Dohlen, K. and Gratton, R. and Boccaletti, A. and Sauvage, J.-F. and Schmid, H. M. and Langlois, M. and Petit, C. and Baruffolo, A. and Feldt, M. and Milli, J. and Wahhaj, Z. and Abe, L. and Anselmi, U. and Antichi, J. and Barette, R. and Baudrand, J. and Baudoz, P. and Bazzon, A. and Bernardi, P. and Blanchard, P. and Brast, R. and Bruno, P. and Buey, T. and Carbillet, M. and Carle, M. and Cascone, E. and Chapron, F. and Charton, J. and Chauvin, G. and Claudi, R. and Costille, A. and De Caprio, V. and De Boer, J. and Delboulbé, A. and Desidera, S. and Dominik, C. and Downing, M. and Dupuis, O. and Fabron, C. and Fantinel, D. and Farisato, G. and Feautrier, P. and Fedrigo, E. and Fusco, T. and Gigan, P. and Ginski, C. and Girard, J. and Giro, E. and Gisler, D. and Gluck, L. and Gry, C. and Henning, T. and Hubin, N. and Hugot, E. and Incorvaia, S. and Jaquet, M. and Kasper, M. and Lagadec, E. and Lagrange, A.-M. and Le Coroller, H. and Le Mignant, D. and Le Ruyet, B. and Lessio, G. and Lizon, J.-L. and Llored, M. and Lundin, L. and Madec, F. and Magnard, Y. and Marteaud, M. and Martinez, P. and Maurel, D. and Ménard, F. and Mesa, D. and Möller-Nilsson, O. and Moulin, T. and Moutou, C. and Origné, A. and Parisot, J. and Pavlov, A. and Perret, D. and Pragt, J. and Puget, P. and Rabou, P. and Ramos, J. and Reess, J.-M. and Rigal, F. and Rochat, S. and Roelfsema, R. and Rousset, G. and Roux, A. and Saisse, M. and Salasnich, B. and Santambrogio, E. and Scuderi, S. and Segransan, D. and Sevin, A. and Siebenmorgen, R. and Soenke, C. and Stadler, E. and Suarez, M. and Tiphène, D. and Turatto, M. and Udry, S. and Vakili, F. and Waters, L. B. F. M. and Weber, L. and Wildi, F. and Zins, G. and Zurlo, A.},
	month = nov,
	year = {2019},
	pages = {A155},
}

@BOOK{Cutri2003,
       author = {{Cutri}, R.~M. and {Skrutskie}, M.~F. and {van Dyk}, S. and {Beichman}, C.~A. and {Carpenter}, J.~M. and {Chester}, T. and {Cambresy}, L. and {Evans}, T. and {Fowler}, J. and {Gizis}, J. and {Howard}, E. and {Huchra}, J. and {Jarrett}, T. and {Kopan}, E.~L. and {Kirkpatrick}, J.~D. and {Light}, R.~M. and {Marsh}, K.~A. and {McCallon}, H. and {Schneider}, S. and {Stiening}, R. and {Sykes}, M. and {Weinberg}, M. and {Wheaton}, W.~A. and {Wheelock}, S. and {Zacarias}, N.},
        title = "{2MASS All Sky Catalog of point sources.}",
         year = 2003,
       adsurl = {https://ui.adsabs.harvard.edu/abs/2003tmc..book.....C},
      publisher = {Infrared Science Archive}
}

@article{marois_angular_2006,
	title = {Angular {Differential} {Imaging}: {A} {Powerful} {High}‐{Contrast} {Imaging} {Technique}},
	volume = {641},
	issn = {0004-637X, 1538-4357},
	shorttitle = {Angular {Differential} {Imaging}},
	url = {https://iopscience.iop.org/article/10.1086/500401},
	doi = {10.1086/500401},
	language = {en},
	number = {1},
	urldate = {2025-11-25},
	journal = {ApJ},
	author = {Marois, Christian and Lafreniere, David and Doyon, Rene and Macintosh, Bruce and Nadeau, Daniel},
	month = apr,
	year = {2006},
	pages = {556--564},
}

@misc{currie_direct_2023,
	url = {http://arxiv.org/abs/2205.05696},
	doi = {10.48550/arXiv.2205.05696},
	language = {en},
	urldate = {2025-11-24},
	publisher = {arXiv},
	author = {Currie, Thayne and Biller, Beth and Lagrange, Anne-Marie and Marois, Christian and Guyon, Olivier and Nielsen, Eric and Bonnefoy, Mickael and Rosa, Robert De},
	month = jul,
	year = {2023},
	note = {{P}rotostars and Planets VII, in press, arXiv:2205.05696},
}

@article{pecaut_star_2016,
	title = {The star formation history and accretion-disc fraction among the {K}-type members of the {Scorpius}–{Centaurus} {OB} association},
	volume = {461},
	issn = {0035-8711, 1365-2966},
	url = {https://academic.oup.com/mnras/article-lookup/doi/10.1093/mnras/stw1300},
	doi = {10.1093/mnras/stw1300},
	language = {en},
	number = {1},
	urldate = {2025-11-24},
	journal = {MNRAS},
	author = {Pecaut, Mark J. and Mamajek, Eric E.},
	month = sep,
	year = {2016},
	pages = {794--815},
}

@article{bohn_young_2020,
	title = {The {Young} {Suns} {Exoplanet} {Survey}: {Detection} of a wide-orbit planetary-mass companion to a solar-type {Sco}-{Cen} member},
	volume = {492},
	copyright = {http://creativecommons.org/licenses/by/4.0/},
	issn = {0035-8711, 1365-2966},
	shorttitle = {The {Young} {Suns} {Exoplanet} {Survey}},
	url = {https://academic.oup.com/mnras/article/492/1/431/5680498},
	doi = {10.1093/mnras/stz3462},
	language = {en},
	number = {1},
	urldate = {2025-11-24},
	journal = {MNRAS},
	author = {Bohn, A J and Kenworthy, M A and Ginski, C and Manara, C F and Pecaut, M J and de Boer, J and Keller, C U and Mamajek, E E and Meshkat, T and Reggiani, M and Todorov, K O and Snik, F},
	month = feb,
	year = {2020},
	pages = {431--443},
}

@article{bohn_two_2020,
	title = {Two {Directly} {Imaged}, {Wide}-orbit {Giant} {Planets} around the {Young}, {Solar} {Analog} {TYC} 8998-760-1$^{\textrm{*}}$},
	volume = {898},
	issn = {2041-8205, 2041-8213},
	url = {https://iopscience.iop.org/article/10.3847/2041-8213/aba27e},
	doi = {10.3847/2041-8213/aba27e},
	language = {en},
	number = {1},
	urldate = {2025-11-24},
	journal = {ApJL},
	author = {Bohn, Alexander J. and Kenworthy, Matthew A. and Ginski, Christian and Rieder, Steven and Mamajek, Eric E. and Meshkat, Tiffany and Pecaut, Mark J. and Reggiani, Maddalena and De Boer, Jozua and Keller, Christoph U. and Snik, Frans and Southworth, John},
	month = jul,
	year = {2020},
	pages = {L16},
}

@article{zhang_13co-rich_2021,
	title = {The {13CO}-rich atmosphere of a young accreting super-{Jupiter}},
	volume = {595},
	issn = {0028-0836, 1476-4687},
	url = {https://www.nature.com/articles/s41586-021-03616-x},
	doi = {10.1038/s41586-021-03616-x},
	language = {en},
	number = {7867},
	urldate = {2025-11-24},
	journal = {Nature},
	author = {Zhang, Yapeng and Snellen, Ignas A. G. and Bohn, Alexander J. and Mollière, Paul and Ginski, Christian and Hoeijmakers, H. Jens and Kenworthy, Matthew A. and Mamajek, Eric E. and Meshkat, Tiffany and Reggiani, Maddalena and Snik, Frans},
	month = jul,
	year = {2021},
	pages = {370--372},
}

@article{zhang_eso_2024,
	title = {The {ESO} {SupJup} {Survey} {III}: {Confirmation} of {13CO} in {YSES} 1 b and {Atmospheric} {Detection} of {YSES} 1 c with {CRIRES}+},
	volume = {168},
	issn = {0004-6256, 1538-3881},
	shorttitle = {The {ESO} {SupJup} {Survey} {III}},
	url = {http://arxiv.org/abs/2409.16660},
	doi = {10.3847/1538-3881/ad7ea9},
	language = {en},
	number = {6},
	urldate = {2025-11-24},
	journal = {AJ},
	author = {Zhang, Yapeng and Picos, Darío González and Regt, Sam de and Snellen, Ignas A. G. and Gandhi, Siddharth and Ginski, Christian and Kesseli, Aurora Y. and Landman, Rico and Mollière, Paul and Nasedkin, Evert and Sánchez-López, Alejandro and Stolker, Tomas and Inglis, Julie and Knutson, Heather A. and Mawet, Dimitri and Wallack, Nicole and Xuan, Jerry W.},
	month = dec,
	year = {2024},
	pages = {246},
}

@article{roberts_new_2025,
	title = {New {Orbital} {Constraints} for {YSES} 1 b and {HR} 2562 {B} from {High}-{Precision} {Astrometry} and {Planetary} {Radial} {Velocities}},
	url = {http://arxiv.org/abs/2509.14321},
	doi = {10.48550/arXiv.2509.14321},
	language = {en},
	urldate = {2025-11-24},
	journal = {AJ},
    volume = {170},
    pages = {273},
	author = {Roberts, Jonathan and Thompson, William and Wang, Jason J. and Blunt, Sarah and Balmer, William O. and Bourdarot, Guillaume and Bowler, Brendan P. and Chauvin, Gael and Eisenhauer, Frank and Henning, Thomas K. and Kammerer, Jens and Kiefer, Flavien and Kenworthy, Matthew A. and Kervella, Pierre and Lacour, Sylvestre and Lagrange, A.-M. and Nielsen, Eric L. and Pueyo, Laurent and Rickman, Emily and Sipilä, Olli and Spezzano, Silvia and Stolker, Tomas and Zurlo, Alice},
	month = sep,
	year = {2025},
}

@article{holstein_survey_2021,
	title = {A survey of the linear polarization of directly imaged exoplanets and brown dwarf companions with {SPHERE}-{IRDIS}. {First} polarimetric detections revealing disks around {DH} {Tau} {B} and {GSC} 6214-210 {B}},
	volume = {647},
	issn = {0004-6361, 1432-0746},
	url = {http://arxiv.org/abs/2101.04033},
	doi = {10.1051/0004-6361/202039290},
	language = {en},
	urldate = {2025-11-24},
	journal = {A\&A},
	author = {Holstein, R. G. van and Stolker, T. and Jensen-Clem, R. and Ginski, C. and Milli, J. and Boer, J. de and Girard, J. H. and Wahhaj, Z. and Bohn, A. J. and Millar-Blanchaer, M. A. and Benisty, M. and Bonnefoy, M. and Chauvin, G. and Dominik, C. and Hinkley, S. and Keller, C. U. and Keppler, M. and Langlois, M. and Marino, S. and Ménard, F. and Perrot, C. and Schmidt, T. O. B. and Vigan, A. and Zurlo, A. and Snik, F.},
	month = mar,
	year = {2021},
	pages = {A21},
}

@article{hoch_silicate_2025,
	title = {Silicate clouds and a circumplanetary disk in the {YSES}-1 exoplanet system},
	volume = {643},
	issn = {0028-0836, 1476-4687},
	url = {https://www.nature.com/articles/s41586-025-09174-w},
	doi = {10.1038/s41586-025-09174-w},
	language = {en},
	number = {8073},
	urldate = {2025-11-24},
	journal = {Nature},
	author = {Hoch, K. K. W. and Rowland, M. and Petrus, S. and Nasedkin, E. and Ingebretsen, C. and Kammerer, J. and Perrin, M. and D’Orazi, V. and Balmer, W. O. and Barman, T. and Bonnefoy, M. and Chauvin, G. and Chen, C. and De Rosa, R. J. and Girard, J. and Gonzales, E. and Kenworthy, M. and Konopacky, Q. M. and Macintosh, B. and Moran, S. E. and Morley, C. V. and Palma-Bifani, P. and Pueyo, L. and Ren, B. and Rickman, E. and Ruffio, J.-B. and Theissen, C. A. and Ward-Duong, K. and Zhang, Y.},
	month = jul,
	year = {2025},
	pages = {938--942},
}

@article{stolker_characterizing_2021,
	title = {Characterizing the protolunar disk of the accreting companion {GQ} {Lupi} {B}},
	volume = {162},
	issn = {0004-6256, 1538-3881},
	url = {http://arxiv.org/abs/2110.04307},
	doi = {10.3847/1538-3881/ac2c7f},
	language = {en},
	number = {6},
	urldate = {2025-11-24},
	journal = {AJ},
	author = {Stolker, Tomas and Haffert, Sebastiaan Y. and Kesseli, Aurora Y. and Holstein, Rob G. van and Aoyama, Yuhiko and Brinchmann, Jarle and Cugno, Gabriele and Girard, Julien H. and Marleau, Gabriel-Dominique and Cugno, Gabriele and Meyer, Michael R. and Milli, Julien and Quanz, Sascha P. and Snellen, Ignas A. G. and Todorov, Kamen O.},
	month = dec,
	year = {2021},
	pages = {286},
}

@article{benisty_circumplanetary_2021,
	title = {A {Circumplanetary} {Disk} around {PDS70c}},
	volume = {916},
	issn = {2041-8205, 2041-8213},
	url = {https://iopscience.iop.org/article/10.3847/2041-8213/ac0f83},
	doi = {10.3847/2041-8213/ac0f83},
	language = {en},
	number = {1},
	urldate = {2025-11-24},
	journal = {ApJL},
	author = {Benisty, Myriam and Bae, Jaehan and Facchini, Stefano and Keppler, Miriam and Teague, Richard and Isella, Andrea and Kurtovic, Nicolas T. and Pérez, Laura M. and Sierra, Anibal and Andrews, Sean M. and Carpenter, John and Czekala, Ian and Dominik, Carsten and Henning, Thomas and Menard, Francois and Pinilla, Paola and Zurlo, Alice},
	month = jul,
	year = {2021},
	pages = {L2},
}

@article{haffert_two_2019,
	title = {Two accreting protoplanets around the young star {PDS} 70},
	volume = {3},
	issn = {2397-3366},
	url = {https://www.nature.com/articles/s41550-019-0780-5},
	doi = {10.1038/s41550-019-0780-5},
	language = {en},
	number = {8},
	urldate = {2025-11-24},
	journal = {NA},
	author = {Haffert, S. Y. and Bohn, A. J. and De Boer, J. and Snellen, I. A. G. and Brinchmann, J. and Girard, J. H. and Keller, C. U. and Bacon, R.},
	month = jun,
	year = {2019},
	pages = {749--754},
}

@article{bailey_thermal_2013,
	title = {A {THERMAL} {INFRARED} {IMAGING} {STUDY} {OF} {VERY} {LOW} {MASS}, {WIDE}-{SEPARATION} {BROWN} {DWARF} {COMPANIONS} {TO} {UPPER} {SCORPIUS} {STARS}: {CONSTRAINING} {CIRCUMSTELLAR} {ENVIRONMENTS}},
	volume = {767},
	copyright = {http://iopscience.iop.org/info/page/text-and-data-mining},
	issn = {0004-637X, 1538-4357},
	shorttitle = {A {THERMAL} {INFRARED} {IMAGING} {STUDY} {OF} {VERY} {LOW} {MASS}, {WIDE}-{SEPARATION} {BROWN} {DWARF} {COMPANIONS} {TO} {UPPER} {SCORPIUS} {STARS}},
	url = {https://iopscience.iop.org/article/10.1088/0004-637X/767/1/31},
	doi = {10.1088/0004-637X/767/1/31},
	language = {en},
	number = {1},
	urldate = {2025-11-24},
	journal = {ApJ},
	author = {Bailey, Vanessa and Hinz, Philip M. and Currie, Thayne and Su, Kate Y. L. and Esposito, Simone and Hill, John M. and Hoffmann, William F. and Jones, Terry and Kim, Jihun and Leisenring, Jarron and Meyer, Michael and Murray-Clay, Ruth and Nelson, Matthew J. and Pinna, Enrico and Puglisi, Alfio and Rieke, George and Rodigas, Timothy and Skemer, Andrew and Skrutskie, Michael F. and Vaitheeswaran, Vidhya and Wilson, John C.},
	month = mar,
	year = {2013},
	pages = {31},
}

@article{bowler_alma_2015,
	title = {{AN} {ALMA} {CONSTRAINT} {ON} {THE} {GSC} 6214-210 {B} {CIRCUM}-{SUBSTELLAR} {ACCRETION} {DISK} {MASS}},
	volume = {805},
	copyright = {http://iopscience.iop.org/info/page/text-and-data-mining},
	issn = {2041-8213},
	url = {https://iopscience.iop.org/article/10.1088/2041-8205/805/2/L17},
	doi = {10.1088/2041-8205/805/2/L17},
	language = {en},
	number = {2},
	urldate = {2025-11-24},
	journal = {ApJ},
	author = {Bowler, Brendan P. and Andrews, Sean M. and Kraus, Adam L. and Ireland, Michael J. and Herczeg, Gregory and Ricci, Luca and Carpenter, John and Brown, Michael E.},
	month = may,
	year = {2015},
	pages = {L17},
}

@article{bowler_disk_2011,
	title = {A {DISK} {AROUND} {THE} {PLANETARY}-{MASS} {COMPANION} {GSC} 06214-00210 b: {CLUES} {ABOUT} {THE} {FORMATION} {OF} {GAS} {GIANTS} {ON} {WIDE} {ORBITS}},
	volume = {743},
	issn = {0004-637X, 1538-4357},
	shorttitle = {A {DISK} {AROUND} {THE} {PLANETARY}-{MASS} {COMPANION} {GSC} 06214-00210 b},
	url = {https://iopscience.iop.org/article/10.1088/0004-637X/743/2/148},
	doi = {10.1088/0004-637X/743/2/148},
	language = {en},
	number = {2},
	urldate = {2025-11-24},
	journal = {ApJ},
	author = {Bowler, Brendan P. and Liu, Michael C. and Kraus, Adam L. and Mann, Andrew W. and Ireland, Michael J.},
	month = dec,
	year = {2011},
	pages = {148},
}

@article{zhou_accretion_2014,
	title = {{ACCRETION} {ONTO} {PLANETARY} {MASS} {COMPANIONS} {OF} {LOW}-{MASS} {YOUNG} {STARS}},
	volume = {783},
	copyright = {http://iopscience.iop.org/info/page/text-and-data-mining},
	issn = {2041-8205, 2041-8213},
	url = {https://iopscience.iop.org/article/10.1088/2041-8205/783/1/L17},
	doi = {10.1088/2041-8205/783/1/L17},
	language = {en},
	number = {1},
	urldate = {2025-11-24},
	journal = {ApJ},
	author = {Zhou, Yifan and Herczeg, Gregory J. and Kraus, Adam L. and Metchev, Stanimir and Cruz, Kelle L.},
	month = feb,
	year = {2014},
	pages = {L17},
}

@article{cugno_carbon-rich_2025,
	title = {A {Carbon}-rich {Disk} {Surrounding} a {Planetary}-mass {Companion}},
	volume = {991},
	issn = {2041-8205, 2041-8213},
	url = {https://iopscience.iop.org/article/10.3847/2041-8213/ae0290},
	doi = {10.3847/2041-8213/ae0290},
	language = {en},
	number = {2},
	urldate = {2025-11-24},
	journal = {ApJL},
	author = {Cugno, Gabriele and Grant, Sierra L.},
	month = oct,
	year = {2025},
	pages = {L46},
}

@article{demars_exoplanet_2025,
	title = {{ExoplaNeT} {accRetion} {mOnitoring} {sPectroscopic} {surveY} ({ENTROPY}) - {II}. {Time} series of {Balmer} line profiles of {Delorme} 1({AB})b},
	copyright = {Creative Commons Attribution 4.0 International},
	url = {https://arxiv.org/abs/2511.01979},
	doi = {10.48550/ARXIV.2511.01979},
	language = {en},
	urldate = {2025-11-24},
	publisher = {arXiv},
	author = {Demars, Dorian and Bonnefoy, Mickaël and Dougados, Catherine and Viswanath, Gayathri and Ringqvist, Simon C. and Janson, Markus and Aoyama, Yuhiko and Thanathibodee, Thanawuth and Marleau, Gabriel-Dominique and Manara, Carlo F. and Rigliaco, Elisabetta and Szulágyi, Judith and Sicilia-Aguilar, Aurora and Bouvier, Jérôme and Alecian, Evelyne and Petrus, Simon and Houllé, Mathis},
	year = {2025},
    journal = {A\&A},
    volume = {706},
    pages = {A57},
}

@article{betti_near-infrared_2022,
	title = {Near-infrared {Accretion} {Signatures} from the {Circumbinary} {Planetary}-mass {Companion} {Delorme} 1 ({AB})b*},
	volume = {935},
	issn = {2041-8205, 2041-8213},
	url = {https://iopscience.iop.org/article/10.3847/2041-8213/ac85ef},
	doi = {10.3847/2041-8213/ac85ef},
	language = {en},
	number = {1},
	urldate = {2025-11-24},
	journal = {ApJL},
	author = {Betti, S. K. and Follette, K. B. and Ward-Duong, K. and Aoyama, Y. and Marleau, G.-D. and Bary, J. and Robinson, C. and Janson, M. and Balmer, W. and Chauvin, G. and Palma-Bifani, P.},
	month = aug,
	year = {2022},
	pages = {L18},
}

@inproceedings{males_magao-x_2024,
	title = {{MagAO}-{X}: {Commissioning} {Results} and {Status} of {Ongoing} {Upgrades}},
	shorttitle = {{MagAO}-{X}},
	url = {http://arxiv.org/abs/2407.13007},
	doi = {10.48550/arXiv.2407.13007},
	language = {en},
	urldate = {2025-11-24},
	publisher = {SPIE},
	author = {Males, Jared R. and Close, Laird M. and Haffert, Sebastiaan Y. and Kautz, Maggie Y. and Kueny, Jay and Long, Joseph D. and McEwen, Eden and Swimmer, Noah and Bailey, John I. and Foster, Warren and Mazin, Benjamin A. and Pearce, Logan and Liberman, Joshua and Twitchell, Katie and Weinberger, Alycia J. and Guyon, Olivier and Hedglen, Alexander D. and McLeod, Avalon and Roberts, Roz and Gorkom, Kyle Van and Li, Jialin and Doty, Isabella and Gasho, Victor},
	month = jul,
	year = {2024},
    volume = {13097},
    booktitle = {Adaptive Optics Systems IX},
    pages = {1309709},
}

@article{amara_span_2012,
	title = {{\textless}span style="font-variant:small-caps;"{\textgreater}pynpoint{\textless}/span{\textgreater} : an image processing package for finding exoplanets: {\textless}span style="font-variant:small-caps;"{\textgreater}pynpoint{\textless}/span{\textgreater}},
	volume = {427},
	issn = {00358711, 13652966},
	shorttitle = {{\textless}span style="font-variant},
	url = {https://academic.oup.com/mnras/article-lookup/doi/10.1111/j.1365-2966.2012.21918.x},
	doi = {10.1111/j.1365-2966.2012.21918.x},
	language = {en},
	number = {2},
	urldate = {2025-11-24},
	journal = {MNRAS},
	author = {Amara, Adam and Quanz, Sascha P.},
	month = dec,
	year = {2012},
	pages = {948--955},
}

@article{long_astrometric_2025,
	title = {Astrometric {Calibration} of {MagAO}-{X} with {Updated} {Solutions} for {HD} 165054 {Field} {Stars}},
	volume = {169},
	issn = {0004-6256, 1538-3881},
	url = {https://iopscience.iop.org/article/10.3847/1538-3881/ad924f},
	doi = {10.3847/1538-3881/ad924f},
	language = {en},
	number = {1},
	urldate = {2025-11-24},
	journal = {AJ},
	author = {Long, Joseph D. and Pearce, Logan and Haffert, Sebastiaan Y. and Males, Jared R. and Close, Laird M. and Guyon, Olivier and Foster, Warren B. and Van Gorkom, Kyle and Hedglen, Alexander D. and Kautz, Maggie Y. and Kueny, Jay K. and Li, Jialin and Lumbres, Jennifer and McEwen, Eden A. and McLeod, Avalon L. and Schatz, Lauren},
	month = jan,
	year = {2025},
	pages = {36},
}

@article{bayo_vosa_2008,
	title = {{VOSA}: virtual observatory {SED} analyzer: {An} application to the {Collinder} 69 open cluster},
	volume = {492},
	issn = {0004-6361, 1432-0746},
	shorttitle = {{VOSA}},
	url = {http://www.aanda.org/10.1051/0004-6361:200810395},
	doi = {10.1051/0004-6361:200810395},
	language = {en},
	number = {1},
	urldate = {2025-11-24},
	journal = {A\&A},
	author = {Bayo, A. and Rodrigo, C. and Barrado Y Navascués, D. and Solano, E. and Gutiérrez, R. and Morales-Calderón, M. and Allard, F.},
	month = dec,
	year = {2008},
	pages = {277--287},
}

@article{stolker_miracles_2020,
	title = {{MIRACLES}: atmospheric characterization of directly imaged planets and substellar companions at 4–5 \textit{μ} m: {I}. {Photometric} analysis of \textit{β} {Pic} b, {HIP} 65426 b, {PZ} {Tel} {B}, and {HD} 206893 {B}},
	volume = {635},
	copyright = {https://www.edpsciences.org/en/authors/copyright-and-licensing},
	issn = {0004-6361, 1432-0746},
	shorttitle = {{MIRACLES}},
	url = {https://www.aanda.org/10.1051/0004-6361/201937159},
	doi = {10.1051/0004-6361/201937159},
	language = {en},
	urldate = {2025-11-24},
	journal = {A\&A},
	author = {Stolker, T. and Quanz, S. P. and Todorov, K. O. and Kühn, J. and Mollière, P. and Meyer, M. R. and Currie, T. and Daemgen, S. and Lavie, B.},
	month = mar,
	year = {2020},
	pages = {A182},
}

@article{gordon_one_2023,
	title = {One {Relation} for {All} {Wavelengths}: {The} {Far}-ultraviolet to {Mid}-infrared {Milky} {Way} {Spectroscopic} {R}({V})-dependent {Dust} {Extinction} {Relationship}},
	volume = {950},
	issn = {0004-637X, 1538-4357},
	shorttitle = {One {Relation} for {All} {Wavelengths}},
	url = {https://iopscience.iop.org/article/10.3847/1538-4357/accb59},
	doi = {10.3847/1538-4357/accb59},
	language = {en},
	number = {2},
	urldate = {2025-11-24},
	journal = {ApJ},
	author = {Gordon, Karl D. and Clayton, Geoffrey C. and Decleir, Marjorie and Fitzpatrick, E. L. and Massa, Derck and Misselt, Karl A. and Tollerud, Erik J.},
	month = jun,
	year = {2023},
	pages = {86},
}

@article{wood_tess_2023,
	title = {{TESS} {Hunt} for {Young} and {Maturing} {Exoplanets} ({THYME}). {IX}. {A} 27 {Myr} {Extended} {Population} of {Lower} {Centaurus} {Crux} with a {Transiting} {Two}-planet {System}},
	volume = {165},
	issn = {0004-6256, 1538-3881},
	url = {https://iopscience.iop.org/article/10.3847/1538-3881/aca8fc},
	doi = {10.3847/1538-3881/aca8fc},
	language = {en},
	number = {3},
	urldate = {2025-11-24},
	journal = {AJ},
	author = {Wood, Mackenna L. and Mann, Andrew W. and Barber, Madyson G. and Bush, Jonathan L. and Kraus, Adam L. and Tofflemire, Benjamin M. and Vanderburg, Andrew and Newton, Elisabeth R. and Feiden, Gregory A. and Zhou, George and Bouma, Luke G. and Quinn, Samuel N. and Armstrong, David J. and Osborn, Ares and Adibekyan, Vardan and Mena, Elisa Delgado and Sousa, Sergio G. and Gagné, Jonathan and Fields, Matthew J. and Milburn, Reilly P. and Thao, Pa Chia and Schmidt, Stephen P. and Gnilka, Crystal L. and Howell, Steve B. and Law, Nicholas M. and Ziegler, Carl and Briceño, César and Ricker, George R. and Vanderspek, Roland and Latham, David W. and Seager, Sara and Winn, Joshua N. and Jenkins, Jon M. and Schlieder, Joshua E. and Osborn, Hugh P. and Twicken, Joseph D. and Ciardi, David R. and Huang, Chelsea X.},
	month = mar,
	year = {2023},
	pages = {85},
}

@inproceedings{allard_model_2011,
	title = {Model {Atmospheres} {From} {Very} {Low} {Mass} {Stars} to {Brown} {Dwarfs}},
	volume = {448},
	language = {en},
	booktitle = {{ASP} {Conference} {Series}},
	publisher = {Astronomical Society of the Pacific},
	author = {Allard, F and Homeier, D and Freytag, B},
	year = {2011},
}

@article{caffau_solar_2011,
	title = {Solar {Chemical} {Abundances} {Determined} with a {CO5BOLD} {3D} {Model} {Atmosphere}},
	volume = {268},
	copyright = {http://www.springer.com/tdm},
	issn = {0038-0938, 1573-093X},
	url = {http://link.springer.com/10.1007/s11207-010-9541-4},
	doi = {10.1007/s11207-010-9541-4},
	language = {en},
	number = {2},
	urldate = {2025-11-24},
	journal = {SP},
	author = {Caffau, E. and Ludwig, H.-G. and Steffen, M. and Freytag, B. and Bonifacio, P.},
	month = feb,
	year = {2011},
	pages = {255--269},
}

@article{allard_bt-settl_2013,
	title = {The {BT}-{Settl} {Model} {Atmospheres} for {Stars}, {Brown} {Dwarfs} and {Planets}},
	volume = {8},
	copyright = {https://www.cambridge.org/core/terms},
	issn = {1743-9213, 1743-9221},
	url = {https://www.cambridge.org/core/product/identifier/S1743921313008545/type/journal_article},
	doi = {10.1017/S1743921313008545},
	language = {en},
	number = {S299},
	urldate = {2025-11-24},
	journal = {Proceedings of the International Astronomical Union},
	author = {Allard, F.},
	month = jun,
	year = {2013},
	pages = {271--272},
}

@article{baraffe_new_2015,
	title = {New evolutionary models for pre-main sequence and main sequence low-mass stars down to the hydrogen-burning limit},
	volume = {577},
	issn = {0004-6361, 1432-0746},
	url = {http://www.aanda.org/10.1051/0004-6361/201425481},
	doi = {10.1051/0004-6361/201425481},
	language = {en},
	urldate = {2025-11-24},
	journal = {A\&A},
	author = {Baraffe, Isabelle and Homeier, Derek and Allard, France and Chabrier, Gilles},
	month = may,
	year = {2015},
	pages = {A42},
}


\clearpage
\begin{appendix}

\vfill\eject

\onecolumn
\FloatBarrier

\section{Posterior distributions for separation, position angle, and contrast for YSES 1 b
\label{mcmc-posterior}}
The posterior distributions of the separation, position angle, and contrast of YSES 1 b in the MagAO-X $i'$-filter and $z'$-filter as estimated by the MCMC analysis.

\begin{figure*}[!h] 
    \begin{subfigure}{0.5\textwidth}
    \centering
    \includegraphics[scale=0.29]{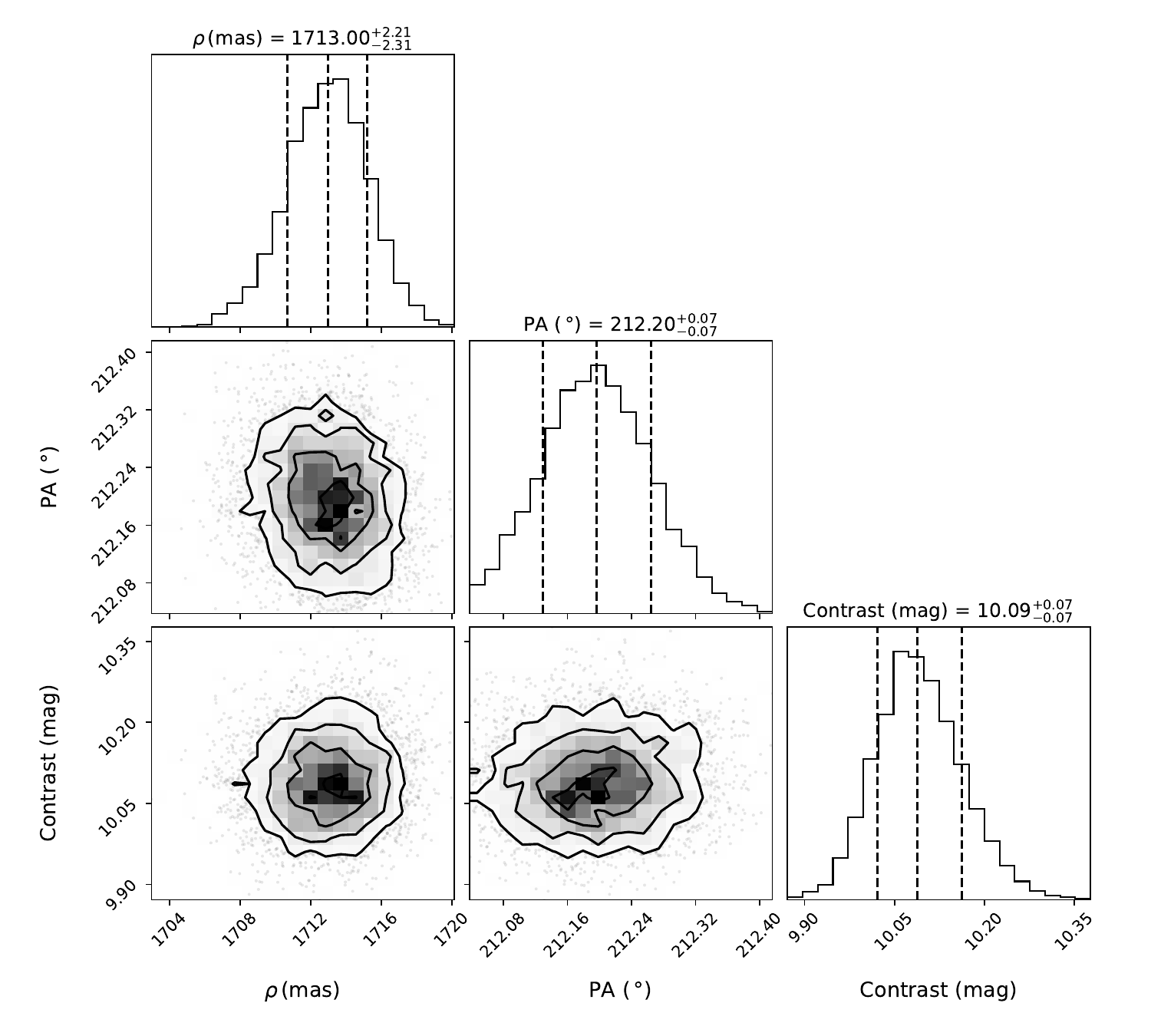}
    \caption{$i'\mathrm{-band}$}
    \end{subfigure}
    \begin{subfigure}{0.5\textwidth}
    \centering
    \includegraphics[scale=0.29]{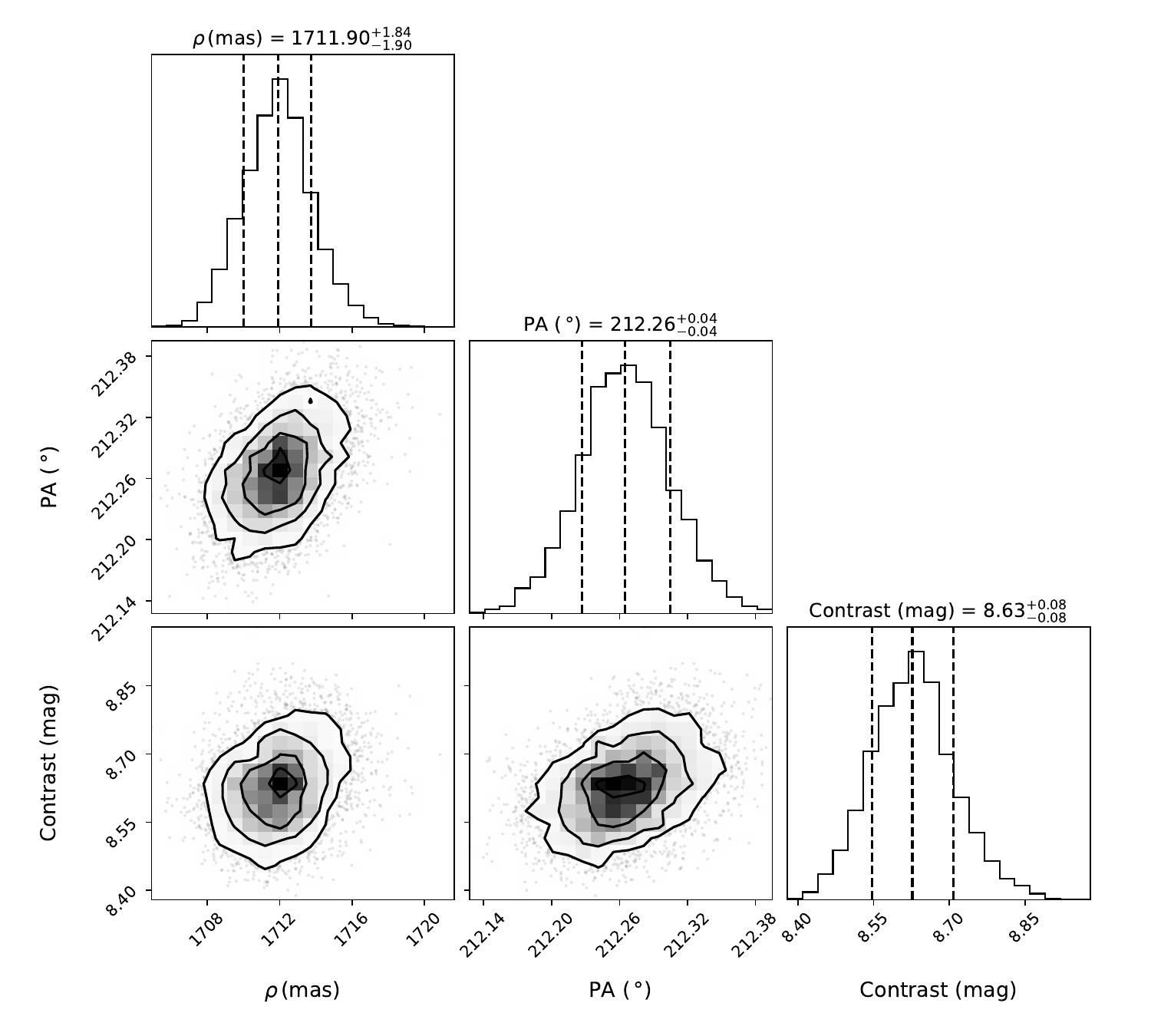} 
    \caption{$z'\mathrm{-band}$}
    \end{subfigure}
    \caption{The posterior distributions for fitting the contrast and astrometry of YSES 1 b in the $i'$-filter and $z'$-filter using the negative artificial planet injection of \texttt{PynPoint} and the \texttt{emcee} sampler \citep{foreman-mackey_emcee_2013}. The plot was made using {\tt corner.py} \citep{foreman-mackey_cornerpy_2016}.}
\end{figure*}



\section{Estimated offsets from systematics analysis}\label{offset}
The offset in separation, position angle, and contrast which is estimated by finding the difference between the injected and retrieved values at different locations.

\begin{figure*}[!h] 
    \begin{subfigure}{\textwidth}
    \centering
    \includegraphics[scale=0.34]{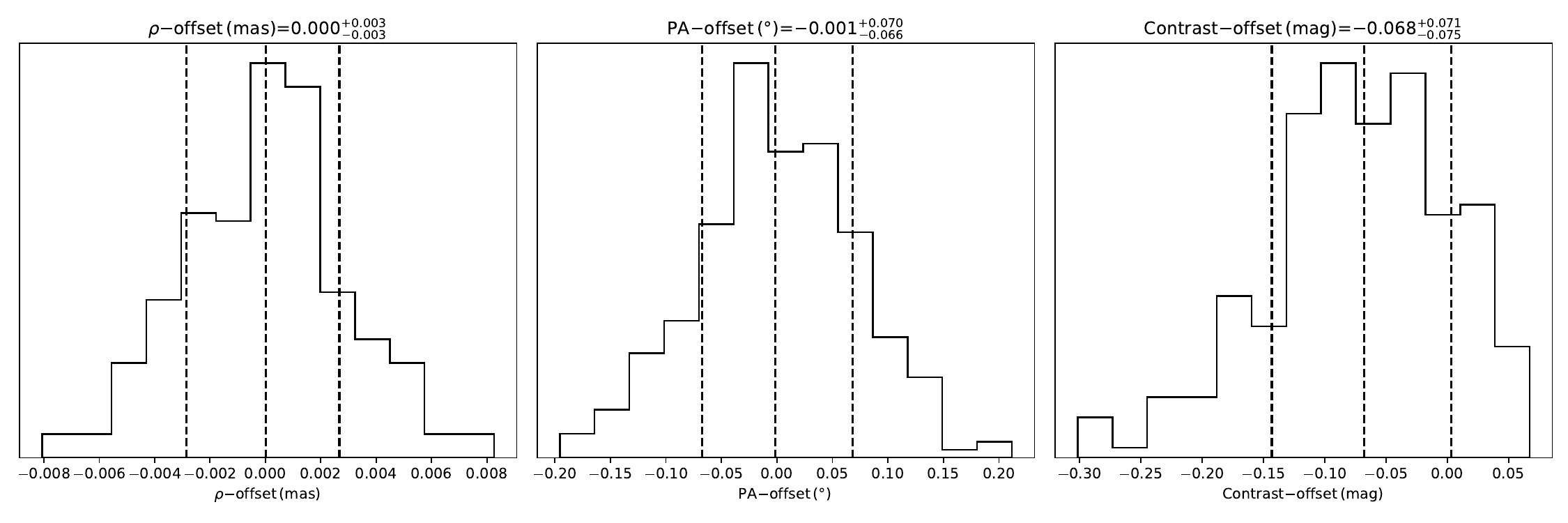}
    \caption{$i'\mathrm{-band}$}
    \end{subfigure}
    \begin{subfigure}{\textwidth}
    \centering
    \includegraphics[scale=0.34]{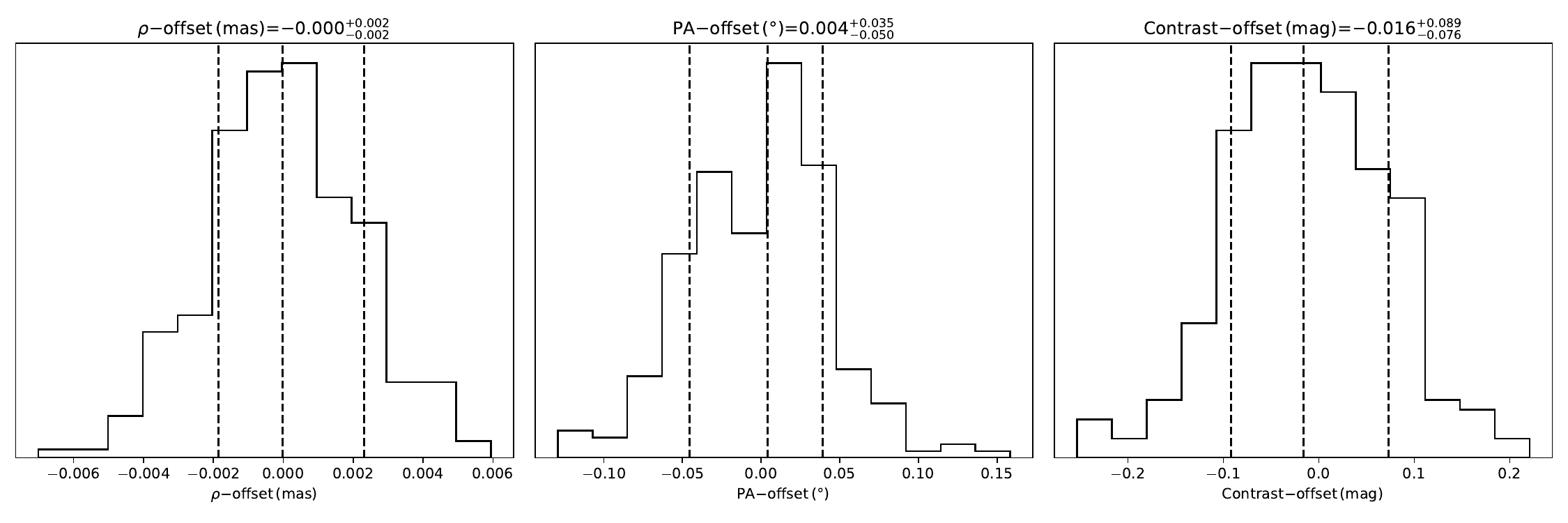} 
    \caption{$z'\mathrm{-band}$}
    \end{subfigure}
    \caption{The estimated distributions for the separation, position angle, and contrast offsets for the $i'$-filter and $z'$-filter. The difference between the injected and retrieved values was evaluated at 250 different position angle locations at the separation of YSES 1 b.}
\end{figure*}

\vfill\eject

\onecolumn
\FloatBarrier

\section{Posterior distributions of the YSES 1 SED fits}\label{stellar-sed-posterior}
Posterior distributions of the stellar SED fit for YSES 1.

\begin{figure*}[!h] 
    \centering
    \includegraphics[scale=0.276]{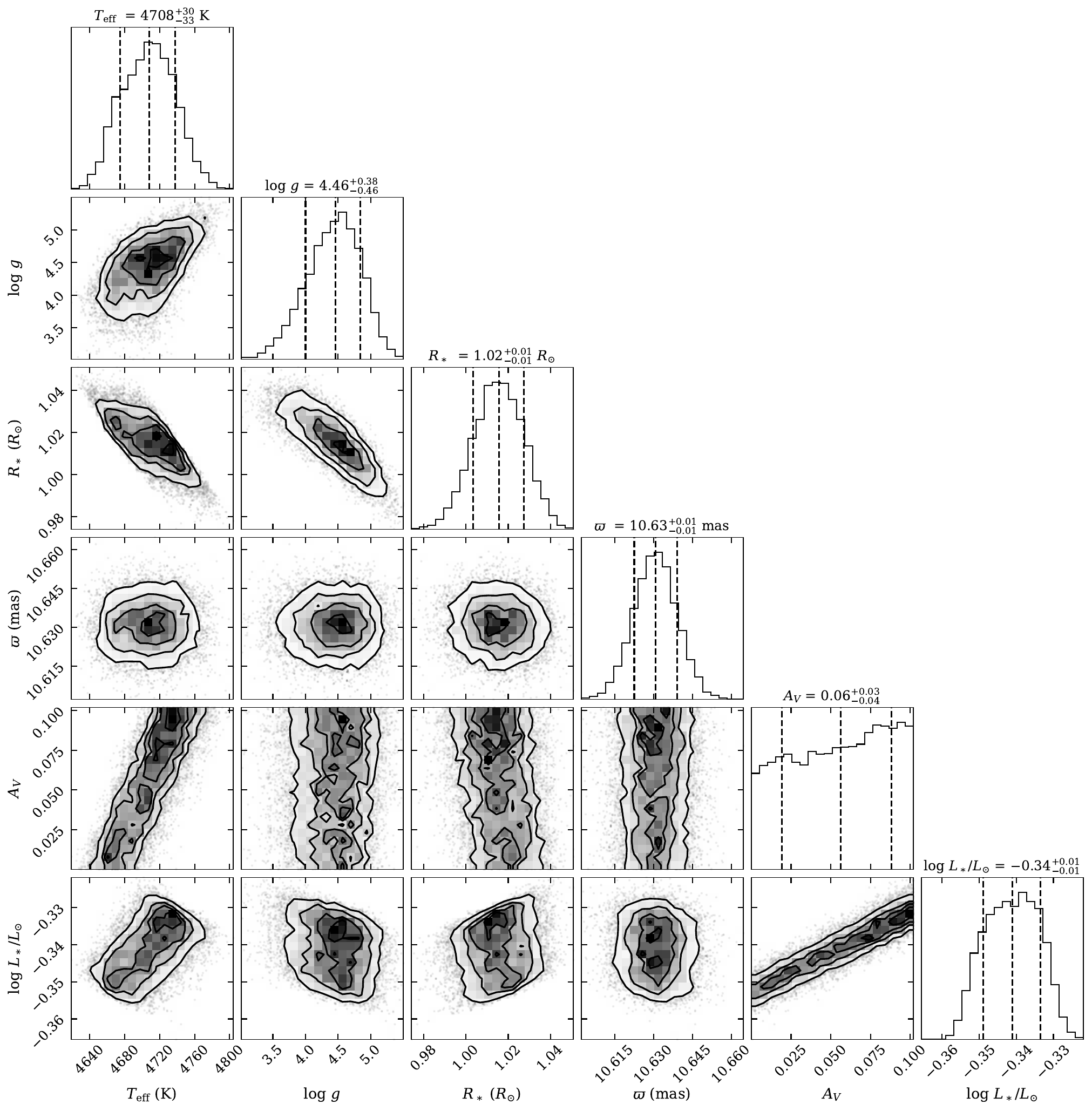}
    \caption{The posterior distributions for the stellar SED fit using the BT-Settl-CIFIST model.
    The nested sampling was performed using \texttt{species} which utilises the \texttt{MultiNest} package \citep{buchner_x-ray_2014}.
    The plot was made using {\tt corner.py} \citep{foreman-mackey_cornerpy_2016}.}
    \label{fig:star-posterior}
\end{figure*}



\section{Posterior distributions of the YSES 1 b SED fits}\label{sed-posterior}
Posterior distributions of the SED fits for the atmospheric model and the atmospheric + CPD model.

\begin{figure*}[!h]
    \centering
    \includegraphics[scale=0.276]{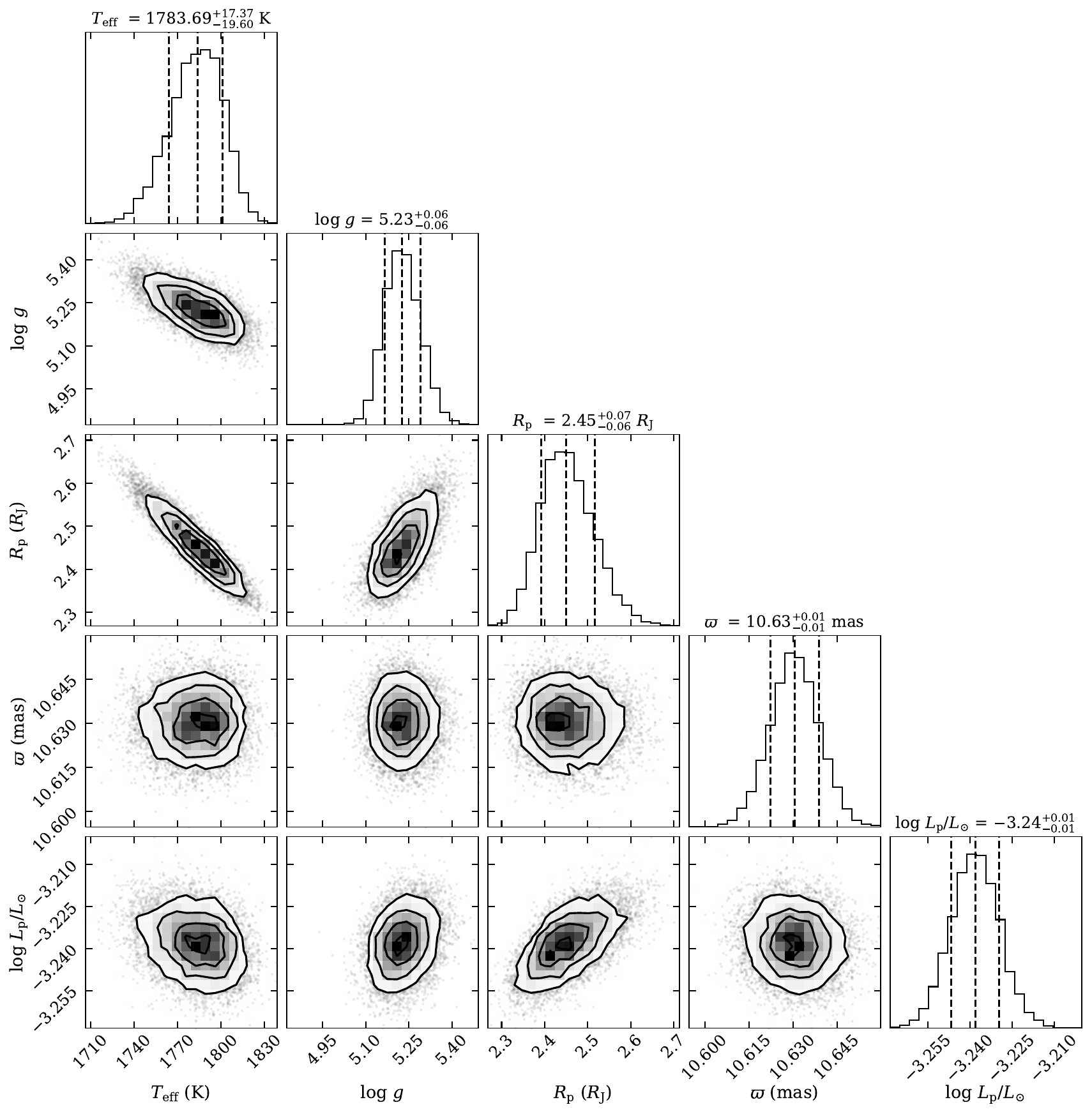}
    \caption{The posterior distributions for fitting the BT-Settl-CIFIST atmospheric model to the photometric data from MagAO-X, SPHERE and NACO.
    The nested sampling was performed using \texttt{species} which utilises the \texttt{MultiNest} package \citep{buchner_x-ray_2014}.
    The plot was made using {\tt corner.py} \citep{foreman-mackey_cornerpy_2016}.}
    \label{fig:no-cpd-posterior}
\end{figure*}

\begin{figure*}[!h] 
    \centering
    \includegraphics[width=\linewidth]{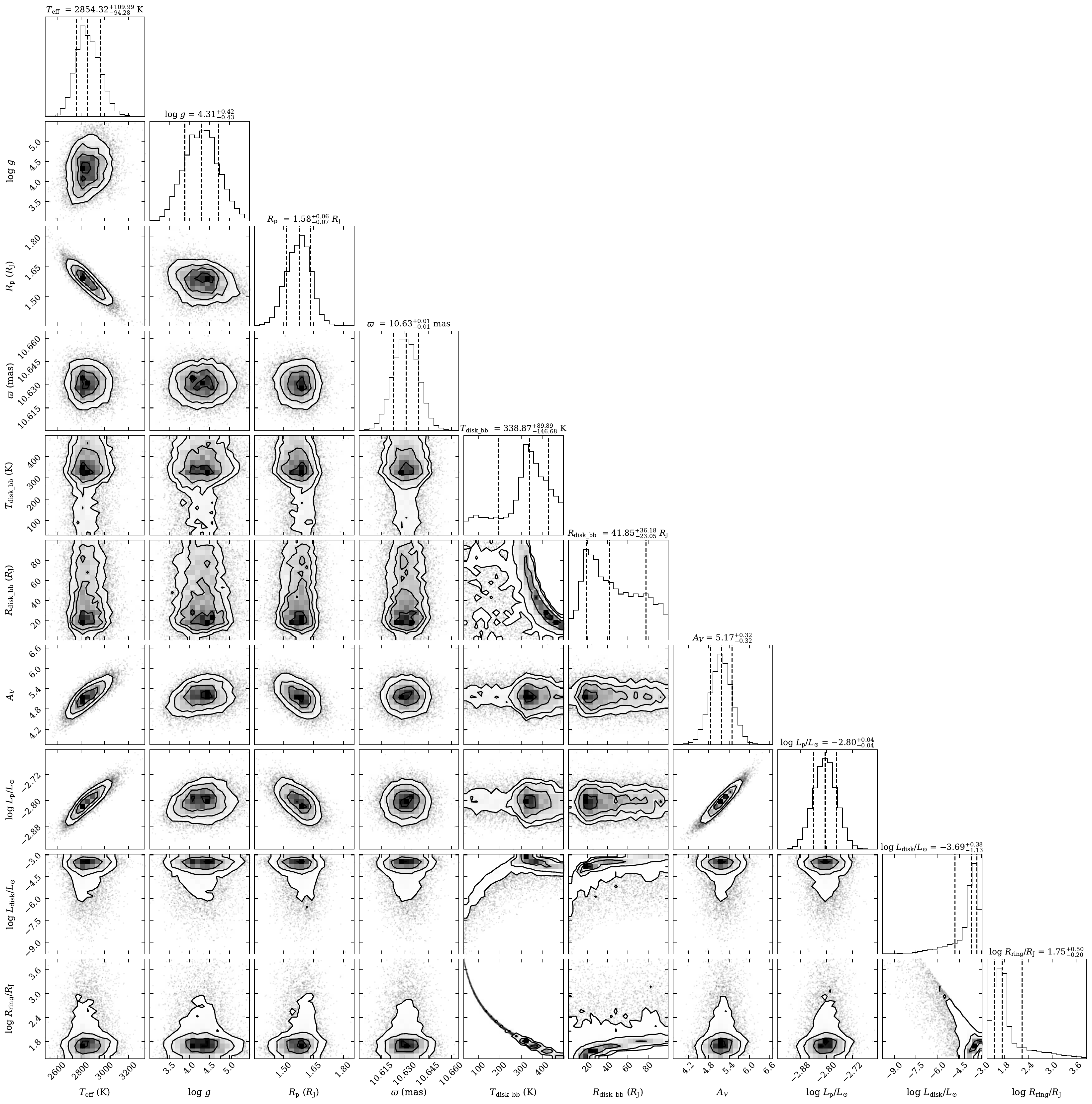}
    \caption{The posterior distributions for fitting the BT-Settl-CIFIST atmospheric model combined with dust extinction and a disc blackbody component to the photometric data from MagAO-X, SPHERE and NACO.
    The nested sampling was performed using \texttt{species} which utilises the \texttt{MultiNest} package \citep{buchner_x-ray_2014}. The plot was made using corner.py \citep{foreman-mackey_cornerpy_2016}.}
    \label{fig:cpd-posterior}
\end{figure*}

\vfill\eject

\onecolumn
\FloatBarrier

\section{Posterior distributions of evolutionary model fits}\label{evolutionary-posterior}
Posterior distributions of the evolutionary model fit for YSES 1 b using the estimated age of 16.7 or 27 Myr.

\begin{figure*}[!h] 
    \begin{subfigure}{\textwidth}
    \centering
    \includegraphics[scale=0.35]{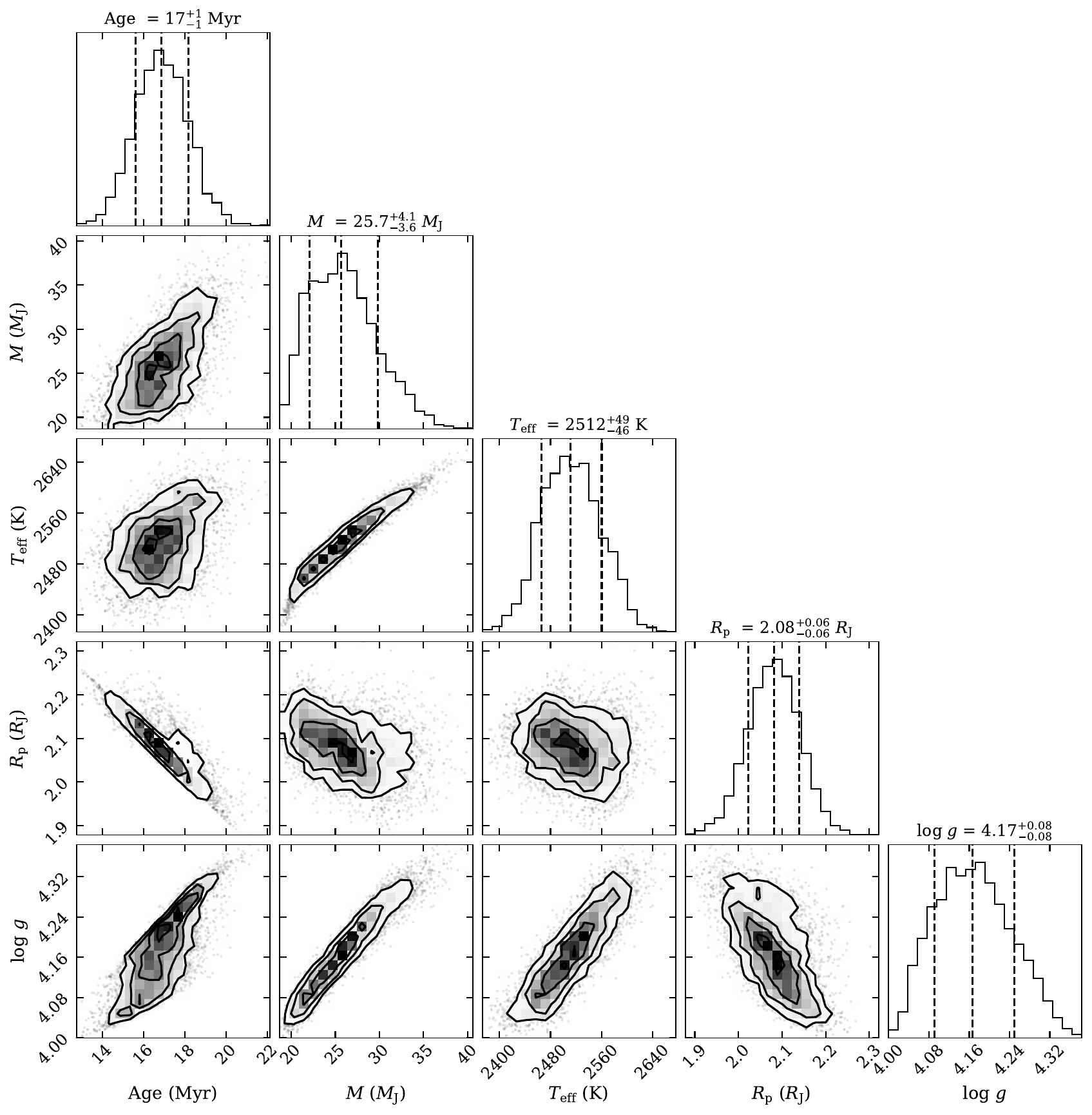}
    \caption{16.7 Myr}
    \end{subfigure}
    \begin{subfigure}{\textwidth}
    \centering
    \includegraphics[scale=0.35]{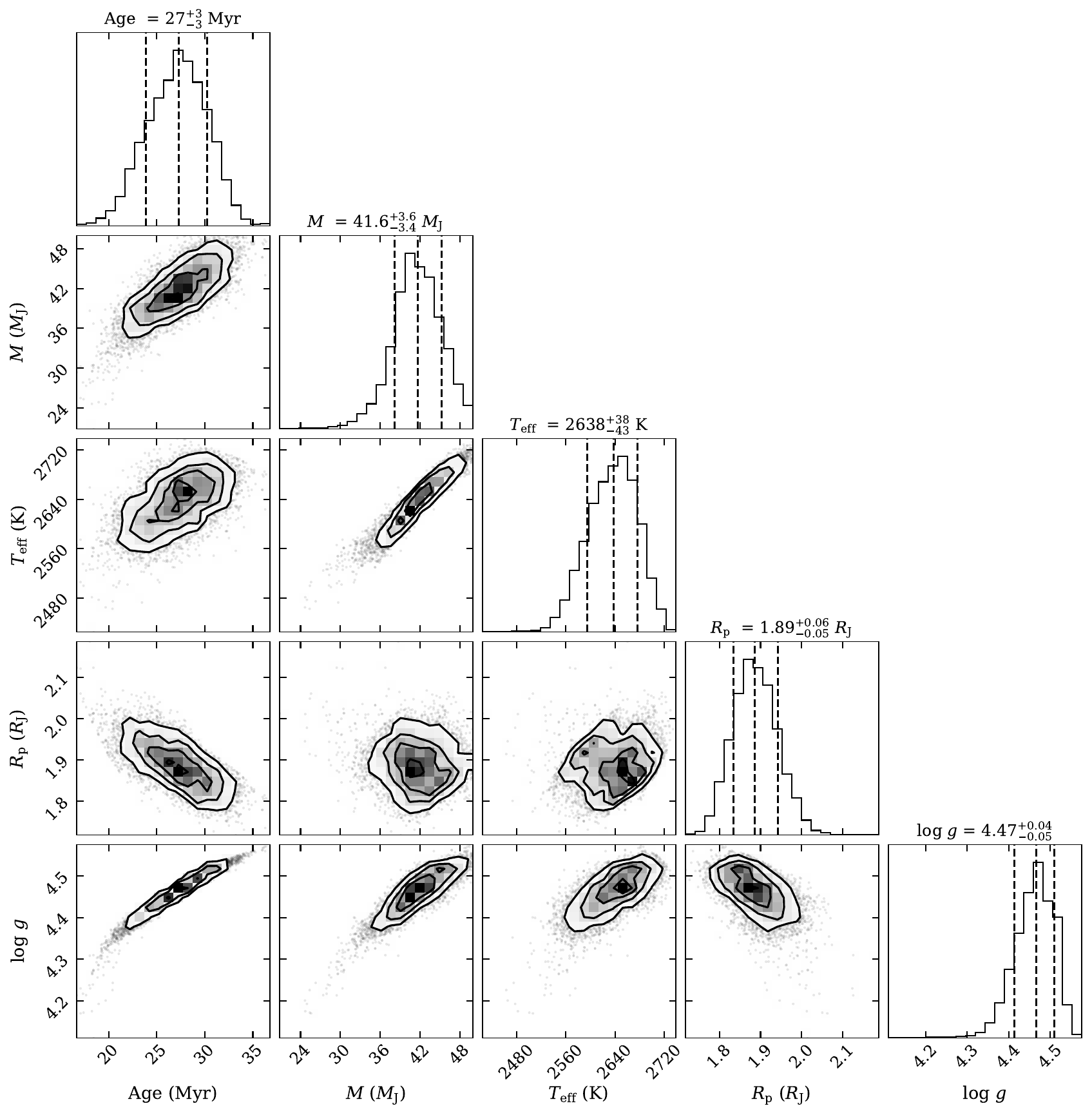}
    \caption{27 Myr}
    \end{subfigure}
    \caption{The posterior distributions for fitting the BT-Settl evolutionary model to the luminosity found in Fig. \ref{fig:cpd-posterior} and assuming a system age of 16.7 Myr or 27 Myr.
    The nested sampling was performed using \texttt{species} which utilises the \texttt{MultiNest} package \citep{buchner_x-ray_2014}.
    The plot was made using {\tt corner.py} \citep{foreman-mackey_cornerpy_2016}.}
\end{figure*}

\end{appendix}

\end{document}